\documentclass[11pt]{article}
\usepackage[a4paper,margin=25mm]{geometry}
\usepackage[T1]{fontenc}
\usepackage{lmodern}
\usepackage{microtype}
\usepackage{amsmath,amssymb,mathtools,bm}
\usepackage{booktabs,array,multirow}
\usepackage{caption}
\usepackage{tabularx}
\usepackage{graphicx}
\usepackage{placeins}
\usepackage{float}
\usepackage{algorithm}
\usepackage{algpseudocode}
\usepackage[backend=biber,style=authoryear-comp,natbib=true,maxcitenames=2,maxbibnames=99,doi=true,isbn=false,url=false]{biblatex}
\usepackage{enumitem}
\usepackage{hyperref}
\usepackage[nameinlink,noabbrev]{cleveref}
\crefname{section}{Section}{Sections}
\Crefname{section}{Section}{Sections}
\crefname{subsection}{Section}{Sections}
\Crefname{subsection}{Section}{Sections}
\crefname{equation}{Equation}{Equations}
\Crefname{equation}{Equation}{Equations}
\crefname{figure}{Figure}{Figures}
\Crefname{figure}{Figure}{Figures}
\crefname{table}{Table}{Tables}
\Crefname{table}{Table}{Tables}
\crefname{algorithm}{Algorithm}{Algorithms}
\Crefname{algorithm}{Algorithm}{Algorithms}
\hypersetup{%
  colorlinks=true,
  linkcolor=blue,
  citecolor=blue,
  urlcolor=blue,
  pdftitle={Plasticity as Directional Stationarity: Yielding, Flow, and Hardening from One Functional},
  pdfauthor={Huilong Ren},
  pdfsubject={Variational formulation of small-strain plasticity},
  pdfkeywords={variational plasticity, directional stationarity, hardening, viscoplasticity}
}
\numberwithin{equation}{section}
\newcommand{\bsig}{\bm\sigma}
\newcommand{\beps}{\bm\varepsilon}
\newcommand{\bepsp}{\bm\varepsilon^{p}}
\newcommand{\balpha}{\bm\alpha}
\newcommand{\bbeta}{\bm\beta}
\newcommand{\bxi}{\bm\xi}
\newcommand{\bM}{\bm M}
\newcommand{\bN}{\bm N}

\newcommand{\dev}{\operatorname{dev}}
\newcommand{\tr}{\operatorname{tr}}
\newcommand{\sym}{\operatorname{sym}}
\newcommand{\dd}{\mathrm d}

\newcommand{\T}{\mathsf T}
\newcommand{\topic}[1]{\par\medskip\noindent\textbf{#1.}\quad}

\title{Plasticity as Directional Stationarity: Yielding, Flow, and Hardening from One Functional}
\author{Huilong Ren\\
\small State Key Laboratory of Disaster Reduction in Civil Engineering,\\
\small College of Civil Engineering, Tongji University, Shanghai 200092, China\\
\small Corresponding author: hlren@tongji.edu.cn}
\date{}

\begin{document}
\maketitle

\begin{abstract}
Traditional plasticity theory is commonly organized through an elastic law, a yield condition, a flow rule, hardening relations, and loading--unloading conditions. This paper formulates these relations through directional stationarity of one scalar functional evaluated over one-sided admissible plastic paths. The first variation determines stress, internal-variable forces, and activity resistance. Restriction to an admissible plastic tangent defines a reduced directional force for each mechanism; its sign and one-sided stationarity give the elastic inequality, complementarity, loading--unloading conditions, and active consistency. Associated response corresponds to alignment between the admissible tangent and the normal to the resulting yield boundary, whereas a nonparallel tangent represents non-associated flow. A self-similarity analysis identifies positively homogeneous stress gauges as a broad associated family and separates the roles of yield-surface shape, isotropic expansion, and kinematic translation. The storage and resistance terms cover isotropic, kinematic, coupled, and gradient hardening, while multiple activity variables describe independently activated mechanisms. A viscous potential extends the construction to rate-dependent evolution. Four closed-form solutions illustrate multi-activity regions, kinematic-hardening fields, pressure-sensitive limit states, and the distinct displacement fields associated with normal and non-normal flow. Time-discrete constitutive integration and dimensional checks are collected in the appendices.
\end{abstract}


\section{Introduction}
\label{sec:ss_introduction}

Plasticity models are usually specified by combining an elastic constitutive law, a yield condition, a plastic flow direction, hardening equations, and loading--unloading rules. Classical work by Hill, Drucker, Moreau, Rice, and the internal-variable thermodynamics literature established the mathematical and physical roles of these ingredients \citep{Hill1950,Drucker1959,Moreau1970,Rice1971,HalphenNguyen1975,Lubliner1990}. This organization remains highly effective in computational plasticity, but it separates relations that share an energetic and directional origin. The present work develops a common variational statement for these relations.

Variational approaches have recently been extended to non-associated flow, gradient plasticity, and thermomechanically coupled dissipation. Non-normal directions can be accommodated by treating the directional data explicitly \citep{Ulloa2021}, while rate-type and incremental principles provide microforce balances and natural boundary terms for gradient and coupled dissipative solids \citep{TeichtmeisterKeip2022}. Recent strain-gradient formulations have also clarified the roles of saturating internal variables, reverse loading, and elastic-gap-free decompositions \citep{AbatourForest2024Saturating,MukherjeeBanerjee2024ElasticGap}. Related data-driven work has emphasized explicit thermodynamic structure for history-dependent constitutive response \citep{FuhgEtAl2025DataDriven}.

The present formulation uses one scalar functional as the generating object and a one-sided tangent set as the admissible variation space. The first variation yields stress, internal-variable forces, and resistance. Its restriction to a plastic tangent defines one reduced directional force for each mechanism, from which the elastic inequality, complementarity, loading--unloading conditions, and active consistency follow. The tangent may be specified directly or represented by a plastic potential; in either case it defines the admissible variation space rather than an additional term in the scalar functional. In the associated case it is selected as a support direction of the admissible plastic metric; a nonparallel tangent describes non-associated response.

Classical yield criteria also reveal a useful scaling structure. The von Mises and Tresca criteria, Hill's anisotropic quadratic form, Hosford's exponent family, the Drucker--Prager pressure-sensitive surface, and Koiter's multisurface construction differ in geometry but are naturally expressed through positively homogeneous stress measures \citep{vonMises1913,Tresca1864,Hill1948Anisotropic,Hosford1972,DruckerPrager1952,Koiter1953}. In the present setting, the stress gauge fixes surface shape and associated direction, the activity resistance fixes its scale, and the plastic-state energy governs translation or other energetic shifts.

This construction differs from established variational frameworks in its choice of primary data. Generalized standard materials combine a free energy with a convex dissipation potential or a dual elastic domain \citep{HalphenNguyen1975}. Hyperplasticity uses energy and dissipation functions as the principal constitutive potentials \citep{CollinsHoulsby1997,HoulsbyPuzrin2007}. Sewell's and Hackl's formulations employ broader generating structures \citep{Sewell1973a,Sewell1973b,Hackl1997,HacklFischer2008}, while Del Piero's energy-stationarity construction derives associated flow and hardening from an energy principle \citep{DelPiero2018Variational}. Here the yield function is identified with the negative first variation per unit admissible plastic activity.

The development below covers mixed hardening, multiple plastic activities, gradient resistance, rate dependence, and tensorial plasticity. Four closed-form examples show how these ingredients affect active regions, internal-force fields, and the separation between pressure-sensitive strength and plastic dilatancy.

The paper is organized as follows. \Cref{sec:ss_variational_foundation} presents the scalar prototype, the multidimensional directional formulation, associated and non-associated flow, and the self-similar associated family. \Cref{sec:obstruction} develops hardening, multiple directions, gradient resistance, and tensorial plasticity. \Cref{sec:ss_analytical_solutions} gives closed-form solutions for multi-activity torsion, a non-uniform bar with kinematic hardening, a Hill-type annulus, and elastic--plastic cavity expansion with independent strength and dilatancy. \Cref{sec:ss_viscous} introduces viscous kinetics and the rate-independent limit, and \Cref{sec:discussion} summarizes the main conclusions. Time-discrete integration and dimensional checks are documented in Appendices~\ref{app:ss_discrete} and~\ref{app:ss_dimensional_check}.

\section{Directional stationarity from one functional}
\label{sec:ss_variational_foundation}

The scalar case is used first to display the calculation with minimal notation.
The same construction is then written in multidimensional stress space, where
association, non-association, corners, and self-similar yield families acquire
their geometric meaning.

\subsection{Scalar prototype and directional stationarity}
\label{subsec:ss_1d_derivation}

Consider a bar with total strain $\varepsilon=u_{,x}$, plastic strain
$\varepsilon^p$, elastic strain $\varepsilon^e=\varepsilon-\varepsilon^p$,
and a nondecreasing activity variable $\lambda$. The scalar functional is
\begin{equation}
  \Pi[u,\varepsilon^p,\lambda]
  =\int_\Omega\left[
    \phi(\varepsilon-\varepsilon^p)
    +\psi^p(\varepsilon^p)
    +W(\lambda)
  \right]\dd x-\ell(u).
  \label{eq:ss_1d_functional_general}
\end{equation}
The corresponding conjugate quantities are
\nopagebreak[4]
\[
  \sigma:=\phi_{,\varepsilon^e},
  \qquad
  \sigma_p:=\psi^p_{,\varepsilon^p},
  \qquad
  R:=W_{,\lambda},
  \qquad
  \xi:=\sigma-\sigma_p.
\]
The first variation is
\begin{align}
  \delta\Pi
  ={}&\int_\Omega\left[
    \phi_{,\varepsilon^e}(\delta\varepsilon-\delta\varepsilon^p)
    +\psi^p_{,\varepsilon^p}\,\delta\varepsilon^p
    +W_{,\lambda}\,\delta\lambda
  \right]\dd x-\delta\ell(u)
  \notag\\
  ={}&\int_\Omega\left[
    \sigma\,\delta\varepsilon
    -\sigma\,\delta\varepsilon^p
    +\sigma_p\,\delta\varepsilon^p
    +R\,\delta\lambda
  \right]\dd x-\delta\ell(u)
  \notag\\
  ={}&\delta\Pi_{\rm eq}
  +\int_\Omega\left[-\xi\,\delta\varepsilon^p
    +R\,\delta\lambda\right]\dd x,
  \label{eq:ss_1d_variation_chain_general}
\end{align}
where $\delta\Pi_{\rm eq}=\int_\Omega\sigma\,\delta\varepsilon\,\dd x-
\delta\ell(u)$. Its stationarity gives the standard equilibrium equation and
traction condition. The remaining term governs plastic evolution.

Parameterize an admissible plastic path by
\begin{equation}
  \delta\varepsilon^p
  =\frac{\partial\varepsilon^p}{\partial\lambda}\,\delta\lambda
  =M\,\delta\lambda,
  \qquad
  M=G_{,\xi}(\xi,\varepsilon^p,\lambda),
  \qquad
  \delta\lambda\ge0.
  \label{eq:ss_1d_general_tangent}
\end{equation}
The plasticity assumption enters through this admissible tangent, which
defines the directional domain of the functional. Directional stationarity
determines the activity along that path. Substitution into
\Cref{eq:ss_1d_variation_chain_general} gives
\begin{align}
  \delta\Pi_{\rm p}
  &=-\int_\Omega F_G\,\delta\lambda\,\dd x,
  \notag\\
  F_G&:=\xi\,G_{,\xi}-R.
  \label{eq:ss_1d_active_relation_direct}
\end{align}
Stability with respect to nonnegative trial variations gives $F_G\le0$, while
stationarity along an active plastic variation gives $F_G\,\delta\lambda=0$.
Hence
\begin{equation}
  \boxed{
  F_G\le0,
  \qquad
  \delta\lambda\ge0,
  \qquad
  F_G\,\delta\lambda=0,
  \qquad
  \delta\varepsilon^p=G_{,\xi}\,\delta\lambda.}
  \label{eq:ss_1d_kkt}
\end{equation}
The quantity $F_G$ is the negative first variation per unit plastic activity;
its zero level is the yield boundary. Together, the functional and its
directional domain define the constitutive model.

For a mixed-hardening prototype, choose
\begin{equation}
  \phi(\varepsilon^e)=\frac12E(\varepsilon^e)^2,
  \qquad
  \psi^p(\varepsilon^p)=\frac12C_p(\varepsilon^p)^2,
  \qquad
  W(\lambda)=\sigma_0\lambda+\frac12H\lambda^2.
  \label{eq:ss_1d_functional}
\end{equation}
Then
\begin{equation}
  \sigma=E(\varepsilon-\varepsilon^p),
  \qquad
  \sigma_p=C_p\varepsilon^p,
  \qquad
  R=\sigma_0+H\lambda,
  \qquad
  \xi=\sigma-\sigma_p.
  \label{eq:ss_1d_forces}
\end{equation}
With $G(\xi)=|\xi|$, the path tangent and yield function are
\begin{equation}
  M=\operatorname{sign}\xi,
  \qquad
  F=|\sigma-\sigma_p|-\sigma_0-H\lambda.
  \label{eq:ss_1d_force_explicit}
\end{equation}
This normalization gives $\delta\lambda=|\delta\varepsilon^p|$. On a smooth
active branch, let $s=\operatorname{sign}\xi$. Variation of $F=0$ gives
\begin{equation}
  0=\delta F
  =sE\,\delta\varepsilon-(E+C_p+H)\,\delta\lambda,
  \label{eq:ss_scalar_delta_F}
\end{equation}
and therefore
\begin{equation}
  \delta\lambda
  =\frac{sE}{E+C_p+H}\,\delta\varepsilon,
  \qquad
  \delta\sigma
  =\frac{E(C_p+H)}{E+C_p+H}\,\delta\varepsilon
  \quad (s\,\delta\varepsilon>0).
  \label{eq:ss_scalar_explicit_multiplier}
\end{equation}
For $s\,\delta\varepsilon\le0$, complementarity gives $\delta\lambda=0$.
Once the admissible tangent has been specified, the reduced variation therefore
determines the yield boundary, loading and unloading, the hardening response,
and the active elastoplastic modulus.

\subsection{Multidimensional functional and admissible directions}
\label{subsec:ss_multidimensional_functional}
\label{sec:notation}
\label{sec:energy}
\label{sec:ss_single_generator}

Internal-variable formulations distinguish state derivatives, which define
thermodynamic forces, from admissible variations, which define irreversible
mechanisms \citep{ColemanNoll1963,Rice1971,Mandel1973,HalphenNguyen1975}. A
leading $\delta$ denotes an infinitesimal variation. In the rate-independent
theory, $\delta\lambda_a\ge0$ is the one-sided activity variation and the
infinitesimal advance along the plastic path. A time parameterization gives
$\delta\lambda_a=\dot\lambda_a\,\delta t$. Dots denote viscous rates and powers,
whereas $\Delta(\cdot)$ denotes a finite increment after time discretization.
Second-order tensors are bold, a colon denotes double contraction, and
$\mathcal X\mathbin{\bullet}\mathcal M$ denotes the duality product of a
generalized force and its conjugate internal-variable direction.

The principal quantities are summarized in \Cref{tab:ss-principal-quantities}.
\begin{table}[H]
\centering
\caption{Principal small-strain quantities. Energy density and stress have the
same physical dimensions.}
\label{tab:ss-principal-quantities}
\begin{tabularx}{\textwidth}{@{}lXl@{}}
\toprule
Symbol & Meaning & Dimension \\
\midrule
$\beps,\bepsp$ & total and plastic infinitesimal strain & $1$ \\
$\bm z_I$ & general internal variable & model dependent \\
$\lambda_a$ & accumulated activity of direction $a$ & $1$ in the examples \\
$\phi,\psi^p,W$ & elastic, plastic-state, and resistance densities & stress \\
$\bsig,\mathcal X_I$ & stress and generalized thermodynamic forces & dual to paired variable \\
$\mathcal M_{Ia}$ & component of direction $a$ in the $\bm z_I$ space & $[\bm z_I]/[\lambda_a]$ \\
$R_a,F_a$ & resistance and reduced directional force & stress \\
$\delta\lambda_a$ & one-sided activity variation & dimension of $\lambda_a$ \\
\bottomrule
\end{tabularx}
\end{table}

At a material state $\bm q$, let the admissible plastic cone be
\begin{equation}
  \mathcal K(\bm q)
  =\left\{(\delta\bm z,\delta\bm\lambda):
  \delta\bm z_I=\sum_{a=1}^{N}\mathcal M_{Ia}(\bm q)\,\delta\lambda_a,
  \ \delta\lambda_a\ge0\right\}.
  \label{eq:ss_admissible_tangent_general}
\end{equation}
The admissible cone may be specified by normalized directions or by an explicit
tangent map $\mathcal M_{Ia}=\partial\bm z_I/\partial\lambda_a$.
Nonnegative activity variations express irreversibility. The relation between
this tangent geometry and the normal to the resulting boundary is examined in
\Cref{subsec:ss_assoc_nonassoc}.

For one plastic mechanism, consider
\begin{equation}
  \Pi_t[\bm u,\bepsp,\lambda]
  =\int_\Omega
  \left[
    \phi(\beps-\bepsp)
    +\psi^p(\bepsp)
    +W(\lambda)
  \right]\dd x
  -\ell_t(\bm u),
  \qquad \beps=\sym\nabla\bm u .
  \label{eq:Pi}
\end{equation}
Here $\phi$ is the elastic energy, $\psi^p$ stores recoverable energy associated
with the plastic state, and $W$ is the activity contribution.  It may contain
both an accumulated rate-independent cost and recoverable hardening storage.  A basic choice is
\begin{equation}
  W(\lambda)=\sigma_0\lambda+\frac12H\lambda^2,
  \qquad R=W_{,\lambda}=\sigma_0+H\lambda .
  \label{eq:ss_basic_W}
\end{equation}
The linear term represents the rate-independent cost and the quadratic term the
recoverable isotropic hardening contribution.

The force derivatives are
\begin{equation}
  \bsig=\phi_{,\beps_e},
  \qquad
  \bsig_p=\psi^p_{,\bepsp},
  \qquad
  R=W_{,\lambda},
  \qquad
  \beps_e=\beps-\bepsp,
  \label{eq:forces}
\end{equation}
with relative plastic force
\begin{equation}
  \bxi=\bsig-\bsig_p.
  \label{eq:ss_relative_force_basic}
\end{equation}
For example,
\begin{equation}
  \phi=\frac12\beps_e:\mathbb C:\beps_e,
  \qquad
  \psi^p=\frac12\bepsp:\mathbb H:\bepsp,
  \qquad
  W=\sigma_0\lambda+H_0(\lambda)
  \label{eq:standard_energy}
\end{equation}
gives $\bsig=\mathbb C:(\beps-\bepsp)$,
$\bsig_p=\mathbb H:\bepsp$, and
$R=\sigma_0+H_{0,\lambda}$.

The first variation is
\begin{align}
  \delta\Pi_t
  ={}&\int_\Omega
  \left[
    \phi_{,\beps_e}:(\delta\beps-\delta\bepsp)
    +\psi^p_{,\bepsp}:\delta\bepsp
    +W_{,\lambda}\,\delta\lambda
  \right]\dd x
  -\delta\ell_t(\bm u)
  \notag\\
  ={}&\int_\Omega\bsig:\delta\beps\,\dd x-
  \delta\ell_t(\bm u)
  +\int_\Omega\left[-\bxi:\delta\bepsp+R\,\delta\lambda\right]\dd x .
  \label{eq:ss_full_first_variation}
\end{align}
Arbitrary displacement variations give weak equilibrium,
\begin{equation}
  \int_\Omega\bsig:\sym\nabla\delta\bm u\,\dd x
  =\delta\ell_t(\bm u).
  \label{eq:ss_weak_equilibrium}
\end{equation}
The internal variation is then restricted by
\begin{equation}
  \delta\bepsp
  =\frac{\partial\bepsp}{\partial\lambda}\,\delta\lambda
  =\bM(\bxi,\bepsp,\lambda)\,\delta\lambda,
  \qquad \delta\lambda\ge0.
  \label{eq:ss_tangent_M}
\end{equation}
Here $\partial\bepsp/\partial\lambda$ is the local tangent to the
admissible plastic path and belongs to the variation space on which the
functional is tested. When convenient, a scalar $G$ parameterizes this tangent through $\bM=G_{,\bxi}$; $G$ defines the admissible direction and is not an additional term in $\Pi_t$.
Substitution gives
\begin{equation}
  \delta\Pi_t
  =\delta\Pi_{\rm eq}
  -\int_\Omega
  \underbrace{\left[\bxi:\bM-R\right]}_{F}
  \delta\lambda\,\dd x,
  \label{eq:first_variation_direction}
\end{equation}
so that
\begin{equation}
  \boxed{F=\bxi:\bM-R.}
  \label{eq:obstruction}
\end{equation}
For several generalized internal variables and independent activities, the
directional restriction becomes
\begin{equation}
  \boxed{F_a=\sum_I\mathcal X_I\mathbin{\bullet}\mathcal M_{Ia}-R_a.}
  \label{eq:shared_obstruction_notation}
\end{equation}
The scalar $F_a$ is the negative first variation per unit activity along the
selected direction.  Its zero level becomes an active boundary after the
one-sided stationarity conditions are imposed.

\subsection{One-sided stationarity and active consistency}
\label{subsec:ss_directional_theorem}

At an equilibrated state, the internal first variation has the local form
$-\bm F\cdot\delta\bm\lambda$.  An admissible one-sided activity variation is
characterized by
\begin{equation}
  \delta\bm\lambda\in\mathbb R_+^N,
  \qquad
  (-\bm F)\cdot(\bm\eta-\delta\bm\lambda)\ge0
  \quad\text{for every }\bm\eta\in\mathbb R_+^N.
  \label{eq:ss_variational_inequality}
\end{equation}
This is directional stationarity on the nonnegative cone.  Taking
$\bm\eta=\delta\bm\lambda+s\bm e_a$ with $s\ge0$ gives $F_a\le0$.
Taking $\bm\eta=\delta\bm\lambda-\delta\lambda_a\bm e_a$, which remains
admissible, gives $F_a\delta\lambda_a\ge0$.  Since
$F_a\le0$ and $\delta\lambda_a\ge0$, the product must vanish.  Thus
\begin{equation}
  \boxed{
  F_a\le0,
  \qquad
  \delta\lambda_a\ge0,
  \qquad
  F_a\,\delta\lambda_a=0.}
  \label{eq:KT}
\end{equation}
The Kuhn--Tucker system is the pointwise form of the one-sided variational
inequality. Under a time parameterization,
$\delta\lambda_a=\dot\lambda_a\,\delta t$ with $\delta t>0$.

Let the internal part of a general first variation be
\begin{equation}
  \delta\Pi_{\rm int}
  =\int_\Omega\left[
    -\sum_{I=1}^{M}\mathcal X_I\mathbin{\bullet}\delta\bm z_I
    +\sum_{a=1}^{N}R_a\,\delta\lambda_a
  \right]\dd x,
  \label{eq:ss_general_internal_variation_theorem}
\end{equation}
where gradient terms have been integrated by parts.  With
\begin{equation}
  \delta\bm z_I
  =\sum_{a=1}^{N}\mathcal M_{Ia}\,\delta\lambda_a,
  \qquad \delta\lambda_a\ge0,
  \label{eq:ss_general_tangent_theorem}
\end{equation}
the derivative along direction $a$ and a nonnegative local test field $\eta_a$
is
\begin{equation}
  \mathrm D\Pi[\bm v_a[\eta_a]]
  =-\int_\Omega F_a\eta_a\,\dd x,
  \qquad
  F_a=\sum_I\mathcal X_I\mathbin{\bullet}\mathcal M_{Ia}-R_a.
  \label{eq:ss_directional_theorem_statement}
\end{equation}
This is the general form of the scalar calculation in
\Cref{subsec:ss_1d_derivation}.

The activity coordinate may be changed without changing the physical path.  If
\begin{equation}
  \widetilde\lambda_a=h_a(\lambda_a),
  \qquad h_a'(\lambda_a)>0,
  \label{eq:ss_general_direction_reparameterization}
\end{equation}
then
\begin{equation}
  \widetilde{\mathcal M}_{Ia}
  =\frac{\mathcal M_{Ia}}{h_a'},
  \qquad
  \widetilde R_a=\frac{R_a}{h_a'},
  \qquad
  \widetilde F_a=\frac{F_a}{h_a'}.
  \label{eq:ss_reparameterized_direction_objects}
\end{equation}
The sign and zero set of $F_a$, the internal-variable variation, and the
associated or non-associated character of a smooth active branch are unchanged.

On a smooth active set $\mathcal A$, decompose an admissible infinitesimal state
change as
$\delta\bm q=\delta\bm q^{\rm ext}+\sum_b\bm v_b\delta\lambda_b$.
Variation of the active equalities gives
\begin{equation}
  0=\delta F_a
  =\delta F_a^{\rm ext}
   +\sum_{b\in\mathcal A}
    \mathcal J^{\rm red}_{ab}\delta\lambda_b,
  \qquad
  \delta F_a^{\rm ext}:=\mathrm D F_a[\delta\bm q^{\rm ext}],
  \qquad
  \mathcal J^{\rm red}_{ab}:=\mathrm D F_a[\bm v_b].
  \label{eq:ss_reduced_jacobian_consistency}
\end{equation}
If the active matrix is nonsingular,
\begin{equation}
  \delta\bm\lambda_{\mathcal A}
  =-\left(\bm J^{\rm red}_{\mathcal A}\right)^{-1}
    \delta\bm F^{\rm ext}_{\mathcal A}.
  \label{eq:ss_reduced_jacobian_multiplier}
\end{equation}
This is the reduced directional Jacobian. It is symmetric for integrable associated directions, whereas state-dependent or non-associated directions generally lead to a nonsymmetric form.

\subsection{Associated and non-associated flow in multiple dimensions}
\label{subsec:ss_assoc_nonassoc}
\label{sec:closures}
\label{subsec:ss_nonsmooth_directions}
\label{sec:ss_representable_scope}

For associated response, the direction can be selected by the first variation.
Let $\mathcal U$ be a compact set of normalized admissible plastic directions
and define its support function
\begin{equation}
  G(\bxi)=\sup_{\bm N\in\mathcal U}\bxi:\bm N.
  \label{eq:ss_support_function}
\end{equation}
Replacing $\mathcal U$ by its convex hull does not change $G$.  At fixed
$\delta\lambda>0$, minimizing the internal first variation is equivalent to
maximizing the directional work,
\begin{equation}
  \bM\in\operatorname*{arg\,max}_{\bm N\in\mathcal U}\bxi:\bm N.
  \label{eq:ss_variational_direction_selection}
\end{equation}
Every maximizing direction satisfies $\bxi:\bM=G(\bxi)$ and belongs to the
subdifferential of the support function; at a nonsmooth point the full
subdifferential is the convex hull of the maximizing directions.  Thus
\begin{equation}
  \bM\in\partial G(\bxi),
  \qquad
  \bxi:\bM=G(\bxi).
  \label{eq:ss_support_subgradient}
\end{equation}
The restricted variation gives
\begin{equation}
  F=G(\bxi)-R.
  \label{eq:ss_support_force}
\end{equation}
If $R$ has no direct dependence on $\bxi$, then at a smooth point
$\bM=G_{,\bxi}=F_{,\bxi}$; normality follows from the support-direction
selection after the plastic metric $\mathcal U$ has been specified.  In one dimension, $\mathcal U=\{-1,+1\}$ and
$G(\xi)=|\xi|$, which is precisely the selection in
\Cref{subsec:ss_1d_derivation}.

The admissible unit set, or equivalently its support function, defines the
plastic metric. Directional minimization selects the active support direction,
and the homogeneous specialization in \Cref{subsec:ss_self_similarity}
produces self-similar associated families. Non-associated response is obtained
by assigning a tangent that is not normal to the resulting boundary.

A regular non-associated direction can be represented locally by a positive
rescaling.  Given a target boundary
$f^{\rm tar}(\bxi,\bm z)=\varphi(\bxi,\bm z)-R(\bm z)$ and a desired direction
$\bm m(\bxi,\bm z)$, assume on a region $\mathcal U$ that
\begin{equation}
  \bxi:\bm m(\bxi,\bm z)\ge c_0>0,
  \qquad
  \varphi(\bxi,\bm z)\ge\varphi_0>0,
  \qquad
  \varphi,\bm m\in C^1(\mathcal U).
  \label{eq:ss_na_regular_region}
\end{equation}
Then
\begin{equation}
  \bM
  =\frac{\varphi(\bxi,\bm z)}{\bxi:\bm m(\bxi,\bm z)}\,\bm m(\bxi,\bm z)
  \label{eq:normalized_nonassoc}
\end{equation}
reproduces
\begin{equation}
  \bxi:\bM-R=f^{\rm tar},
  \label{eq:prescribed_yield}
\end{equation}
provided the rescaled direction remains bounded and continuous.  This
representation is regular on regions where the directional work remains
bounded away from zero and retains a fixed sign.

For thermodynamic interpretation, split the resistance into a dissipative and
a recoverable part,
\begin{equation}
  R=R^{\rm d}+R^{\rm e},
  \qquad R^{\rm e}=H_{,\lambda},
  \qquad R^{\rm d}\ge0.
  \label{eq:ss_resistance_split}
\end{equation}
On an active branch, $\bxi:\bM=R$.  The dissipated work associated with an
admissible plastic variation is therefore
\begin{equation}
  \delta\mathcal W^{\rm d}
  =\left(\bxi:\bM-R^{\rm e}\right)\delta\lambda
  =R^{\rm d}\delta\lambda\ge0.
  \label{eq:ss_na_dissipation_positive}
\end{equation}
At an apex or corner, the admissible tangent is represented by a closed
set-valued graph.  Continuity or monotonicity of this graph together with a
nonsingular active Jacobian provides the local conditions for a unique active
update.  The non-associated consistent tangent is generally nonsymmetric, and
its stability is governed by the active Jacobian.  The normalization in
\Cref{eq:normalized_nonassoc} describes smooth work-positive branches, while
the set-valued construction below covers apexes and corners.

Let $G(\bxi)$ be convex, nonnegative, and positively homogeneous of degree one,
but possibly nonsmooth.  An associated plastic direction may be selected from
its convex subdifferential,
\begin{equation}
  \bM\in\partial G(\bxi),
  \qquad
  G(\bm y)\ge G(\bxi)+\bM:(\bm y-\bxi)
  \quad\text{for all }\bm y.
  \label{eq:ss_subgradient_definition}
\end{equation}
Degree-one homogeneity implies
\begin{equation}
  \bxi:\bM=G(\bxi)
  \qquad\text{for every }\bM\in\partial G(\bxi),
  \label{eq:ss_subgradient_euler}
\end{equation}
and hence
\begin{equation}
  F=G(\bxi)-R,
  \qquad
  \delta\bepsp=\bM\,\delta\lambda,
  \qquad
  \bM\in\partial G(\bxi).
  \label{eq:ss_nonsmooth_directional_structure}
\end{equation}
On a smooth face the subdifferential contains one normal; at an edge or corner
it contains their convex hull.

For
\begin{equation}
  G(\bxi)=\max_{\alpha=1,\ldots,m}g_\alpha(\bxi),
  \label{eq:ss_max_gauge}
\end{equation}
with degree-one smooth gauges $g_\alpha$,
\begin{equation}
  \partial G(\bxi)
  =\operatorname{co}\left\{
    g_{\alpha,\bxi}(\bxi):\alpha\in\mathcal I(\bxi)
  \right\}.
  \label{eq:ss_max_gauge_subdifferential}
\end{equation}
A single activity with a convex-combination direction describes corner flow in
the normal cone.  Separate nonnegative activities are preferable when the
faces can activate independently.

For a signed resolved force $\tau_a$, forward and reverse slip may likewise be
represented by two one-sided directions,
\begin{equation}
  F_a^+=\tau_a-R_a^+,
  \qquad
  F_a^-=-\tau_a-R_a^-.
  \label{eq:ss_forward_reverse_slip}
\end{equation}
The absolute-value form $|\tau_a|-R_a$ is recovered when the two resistances and
hardening structures coincide.  A nonsmooth non-associated model uses a
different single-valued or set-valued admissible map; association then means
inclusion in the normal cone of the resulting boundary.

For a prescribed target boundary $f_a^{\rm tar}$ and target direction
$\bm m_a$, local representability reduces to the compatibility condition
\begin{equation}
  f_a^{\rm tar}+R_a
  =\mathcal X\mathbin{\bullet}\mathcal M_a.
  \label{eq:ss_inverse_design_compatibility}
\end{equation}
For $\mathcal X\mathbin{\bullet}\bm m_a>0$, the rescaling in
\Cref{eq:normalized_nonassoc} gives the required direction.  Regions with
vanishing or sign-changing directional work are represented through an
enlarged state space, additional internal variables, a nonsmooth graph, or a
bipotential construction.  The resulting class includes associated
homogeneous gauges, independent activities, mixed hardening, gradient
resistance, and regular non-associated directions.  Softening, nonconvex
storage, and localization add the corresponding well-posedness and
regularization conditions.

\subsection{Self-similar associated families}
\label{subsec:ss_self_similarity}

The multidimensional associated condition in
\Cref{subsec:ss_assoc_nonassoc} requires the admissible tangent to be parallel
to the normal of the directional yield boundary. Consider the constant
proportionality
\begin{equation}
  G_{,\bxi}=kF_{,\bxi},
  \qquad k>0,
  \label{eq:ss_selfsimilar_alignment}
\end{equation}
with a resistance $R(\bm z)$ that has no direct dependence on $\bxi$.
Substitution of \Cref{eq:ss_selfsimilar_alignment} into
$F=\bxi:G_{,\bxi}-R$ gives
\begin{equation}
  F=k\,\bxi:F_{,\bxi}-R.
  \label{eq:ss_selfsimilar_pde}
\end{equation}
Let $\Phi:=F+R$ and write $\bxi=\varrho\bm n$, where $\varrho>0$ and $\bm n$
is fixed along a stress ray. Because $R$ is constant on that ray,
\Cref{eq:ss_selfsimilar_pde} becomes
\begin{equation}
  k\varrho\frac{\partial\Phi}{\partial\varrho}=\Phi,
  \qquad
  \Phi(\varrho\bm n,\bm z)
  =\varrho^{1/k}\Phi(\bm n,\bm z).
  \label{eq:ss_selfsimilar_ray_solution}
\end{equation}
Hence, for a reference stress $\sigma_*>0$,
\begin{equation}
  F(\bxi,\bm z)
  =\sigma_*\widehat\Phi_m(\bxi/\sigma_*,\bm z)-R(\bm z),
  \qquad
  \widehat\Phi_m(a\widehat{\bxi},\bm z)
  =a^m\widehat\Phi_m(\widehat{\bxi},\bm z),
  \qquad
  m=\frac1k,
  \label{eq:ss_selfsimilar_general_solution}
\end{equation}
where $\widehat\Phi_m$ is dimensionless and positively homogeneous of degree
$m$. The factor $\sigma_*$ preserves the stress dimension of $F$ for every
$m$. The directional stationarity equation therefore defines a self-similar
family rather than a single yield function.

A convenient parametrization starts from a positive degree-one stress gauge
$\rho(\bxi,\bm z)$. Set
\begin{equation}
  G_m(\bxi,\bm z)
  =\frac{\sigma_*}{m}\left(\frac{\rho}{\sigma_*}\right)^m,
  \qquad
  \bM_m=G_{m,\bxi}
  =\left(\frac{\rho}{\sigma_*}\right)^{m-1}\rho_{,\bxi}.
  \label{eq:ss_selfsimilar_gauge_potential}
\end{equation}
Euler's identity $\bxi:\rho_{,\bxi}=\rho$ gives
\begin{equation}
  \bxi:\bM_m
  =\sigma_*\left(\frac{\rho}{\sigma_*}\right)^m,
  \qquad
  F_m
  =\sigma_*\left(\frac{\rho}{\sigma_*}\right)^m-R,
  \qquad
  F_{m,\bxi}=m\bM_m.
  \label{eq:ss_selfsimilar_gauge_family}
\end{equation}
At smooth points every member is associated; at corners the statement holds
in the subdifferential sense. For the elementary resistance
$W(\lambda)=\sigma_0\lambda+H\lambda^2/2$, representative members are
\begin{equation}
  F_1=\rho-(\sigma_0+H\lambda),\qquad
  F_2=\frac{\rho^2}{\sigma_*}-(\sigma_0+H\lambda),\qquad
  F_4=\frac{\rho^4}{\sigma_*^3}-(\sigma_0+H\lambda).
  \label{eq:ss_selfsimilar_representative_members}
\end{equation}
More generally, for $R>0$, the yield surface satisfies
\begin{equation}
  \mathcal Y(\bm z)
  =\left\{\bxi:\rho(\bxi,\bm z)
  =\sigma_*\left(\frac{R(\bm z)}{\sigma_*}\right)^{1/m}\right\}.
  \label{eq:ss_selfsimilar_surface_scaling}
\end{equation}
Thus a single gauge generates a family of geometrically similar surfaces, while
the exponent changes the relation between accumulated activity and radial
expansion. In stress space, $\bxi=\bsig-\bsig_p$ translates this homothetic
family by the backstress $\bsig_p$. The gauge $\rho$ controls shape and flow
direction, $W$ controls radial scale through $R=W_{,\lambda}$, and $\psi^p$
controls the center or other energetic internal forces. State dependence of
$\rho$ extends the construction to evolving anisotropy or distortional
hardening.

\begin{figure}[t]
  \centering
  \includegraphics[width=0.93\textwidth]{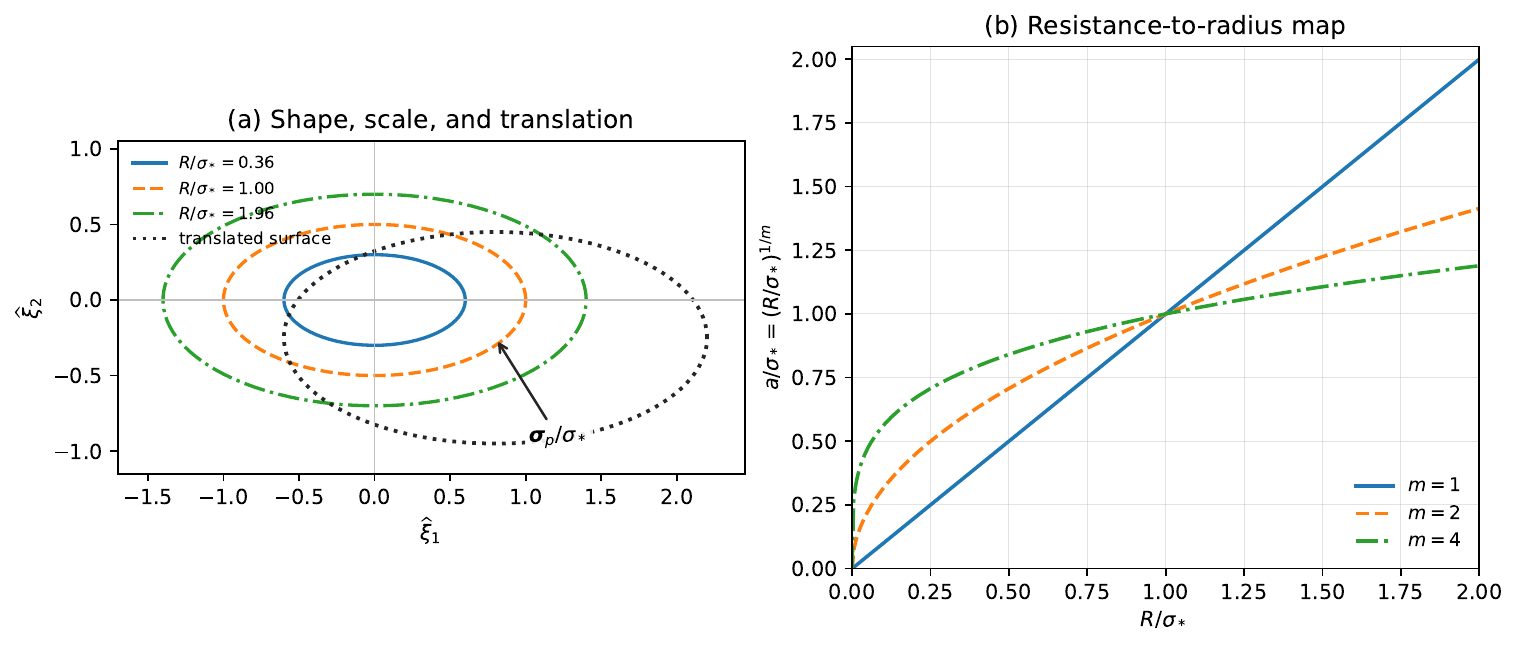}
  \caption{Self-similar directional families. The left panel uses
  $\rho_{\rm H}/\sigma_*=(\widehat{\xi}_1^2+4\widehat{\xi}_2^2)^{1/2}$,
  $m=2$, and $R/\sigma_*=0.36,1.00,1.96$; the dotted contour translates the
  largest surface by $\bsig_p/\sigma_*=(0.8,-0.25)$. The right panel shows
  the normalized radial scale $a/\sigma_*=(R/\sigma_*)^{1/m}$ for
  $m=1,2,4$. The gauge fixes the shape, $R$ fixes the
  size, and the energetic backstress fixes the center.}
  \label{fig:ss_self_similarity}
\end{figure}

Several classical criteria follow as direct choices of the gauge. With a
symmetric positive semidefinite fourth-order tensor $\mathbb H$,
\begin{equation}
  \rho_{\rm H}(\bxi)=\sqrt{\bxi:\mathbb H:\bxi},
  \qquad
  F_m^{\rm H}
  =\sigma_*^{1-m}(\bxi:\mathbb H:\bxi)^{m/2}-R,
  \qquad
  \bM_m^{\rm H}
  =\sigma_*^{1-m}(\bxi:\mathbb H:\bxi)^{m/2-1}\mathbb H:\bxi.
  \label{eq:ss_selfsimilar_hill_family}
\end{equation}
The cases $m=1$ and $m=2$ give the square-root and quadratic Hill forms,
respectively; isotropic specialization recovers the von Mises family
\citep{vonMises1913,Hill1948Anisotropic}. A pressure-sensitive gauge
$\rho_{\rm p}=q+\alpha p$, with
$p=\tr\bxi/3$ and $q=\sqrt{3\dev\bxi:\dev\bxi/2}$, gives
\begin{equation}
  F_m^{\rm p}=\sigma_*^{1-m}(q+\alpha p)^m-R,
  \qquad
  \bM_m^{\rm p}
  =\left(\frac{q+\alpha p}{\sigma_*}\right)^{m-1}
  \left(\frac{3\dev\bxi}{2q}+\frac{\alpha}{3}\bm I\right),
  \label{eq:ss_selfsimilar_pressure_family}
\end{equation}
for $q+\alpha p>0$, including the associated Drucker--Prager form when $m=1$
\citep{DruckerPrager1952}. A polyhedral gauge gives
\begin{equation}
  \rho_{\rm poly}(\bxi)
  =\max_{\alpha}|\bm L_\alpha:\bxi|,
  \qquad
  F_m^{\rm poly}
  =\sigma_*\left(\frac{\rho_{\rm poly}}{\sigma_*}\right)^m-R,
  \qquad
  \bM_m^{\rm poly}
  \in\left(\frac{\rho_{\rm poly}}{\sigma_*}\right)^{m-1}
  \partial\rho_{\rm poly},
  \label{eq:ss_selfsimilar_polyhedral_family}
\end{equation}
which recovers Tresca- and Koiter-type corners and multisurface flow through a
set-valued tangent \citep{Tresca1864,Koiter1953,Moreau1970}. Hosford-type and
other smooth anisotropic criteria correspond to alternative degree-one gauges
in \Cref{eq:ss_selfsimilar_gauge_family}
\citep{Hosford1972,Barlat1991,CazacuBarlat2004}.
\Cref{fig:ss_self_similarity} illustrates the resulting separation of shape,
radial scale, and kinematic translation.

\section{Hardening, multiple directions, gradients, and tensorial plasticity}
\label{sec:obstruction}

Let \(\bm\lambda=(\lambda_1,\ldots,\lambda_N)\) and allow
\begin{equation}
  \Pi_t
  =\int_\Omega
  \left[
    \phi(\beps-\bepsp)
    +\psi^p(\bepsp,\bm z)
    +W(\bm\lambda,\nabla\bm\lambda)
  \right]\dd x-\ell_t(\bm u).
  \label{eq:ss_multilambda_functional}
\end{equation}
Admissible variations take the form
\begin{equation}
  \delta\bepsp
  =\sum_{a=1}^{N}\bm N_a\delta\lambda_a,
  \qquad
  \delta\bm z_I=\sum_{a=1}^{N}\mathcal M_{Ia}\delta\lambda_a,
  \qquad
  \delta\lambda_a\ge0.
  \label{eq:ss_multidirection_variation}
\end{equation}
After integration by parts, the effective resistance is
\begin{equation}
  R_a
  =W_{,\lambda_a}
   -\operatorname{div}W_{,\nabla\lambda_a},
  \label{eq:ss_gradient_resistance}
\end{equation}
with natural micro-boundary condition
\(W_{,\nabla\lambda_a}\cdot\bm n=0\).  The directional variation becomes
\begin{equation}
  \delta\Pi_{\rm int}
  =-\int_\Omega\sum_{a=1}^{N}F_a\delta\lambda_a\,\dd x,
  \qquad
  F_a
  =\bxi:\bm N_a
   +\sum_I\mathcal X_I\mathbin{\bullet}\mathcal M_{Ia}
   -R_a .
  \label{eq:first_variation_direction_multi}
\end{equation}
For compatibility with later notation, this general result is also written as
\begin{equation}
  F_a=\sum_I\mathcal X_I\mathbin{\bullet}\mathcal M_{Ia}-R_a,
  \label{eq:obstruction_multi}
\end{equation}
where the plastic-strain force and direction are included among the generalized
pairs.  Each independent \(\delta\lambda_a\) generates its own inequality,
\begin{equation}
  F_a\le0,
  \qquad
  \delta\lambda_a\ge0,
  \qquad
  F_a\,\delta\lambda_a=0.
  \label{eq:admissibility_multi}
\end{equation}
Coupling in \(W\) produces coupled or latent hardening, gradients produce
nonlocal resistance, and distinct \(\lambda_a\) provide independently
activatable mechanisms. Isotropic hardening, surface translation, latent interaction, and gradient
resistance thus enter through distinct derivatives of the scalar functional,
with the active-set structure used in multisurface and crystal-plasticity
models
\citep{Koiter1953,Iwan1967,Asaro1983,PeirceAsaroNeedleman1983}.

\subsection{Dissipation}

Write
\(W=\sum_a\sigma_{0a}\lambda_a+H(\bm\lambda)\), with \(H\) recoverable.
The energetic hardening force is \(H_{,\lambda_a}\), while the linear term is
the rate-independent direction cost.  On an active direction,
$F_a=0$ gives
\begin{equation}
  \mathcal X\mathbin{\bullet}\mathcal M_a-H_{,\lambda_a}
  =\sigma_{0a},
\end{equation}
where the generalized product includes all internal-variable pairs.  Hence the dissipated work over an admissible plastic variation is
\begin{equation}
  \delta\mathcal W^{\rm d}
  =\sum_a
  \left(\mathcal X\mathbin{\bullet}\mathcal M_a-H_{,\lambda_a}\right)
  \delta\lambda_a
  =\sum_a\sigma_{0a}\delta\lambda_a\ge0.
  \label{eq:dissipation}
\end{equation}
A physical dissipation rate is obtained by the optional parametrization
$\delta\lambda_a=\dot\lambda_a\,\delta t$.  Recovery and viscous contributions enter through their corresponding
directional work and rate terms below.

\subsection{Hardening and the role of multiple accumulated variables}

The following examples summarize linear and nonlinear kinematic hardening,
cyclic memory, and ratcheting models developed from Prager through modern
multi-backstress formulations \citep{Prager1956,ArmstrongFrederick1966,
Mroz1967,DafaliasPopov1976,Chaboche1986,Chaboche1989,OhnoWang1993,
AbdelKarimOhno2000}.

\label{sec:multilambda}

\topic{Isotropic, kinematic, and coupled hardening}

With the energy \Cref{eq:standard_energy}, the resistance vector is
\begin{equation}
  R_a=\sigma_{0a}+\frac{\partial H}{\partial\lambda_a}.
\end{equation}
For quadratic coupled hardening, take a symmetric matrix $\bm H$ (positive
semidefinite when convex hardening is required),
\begin{equation}
  H(\bm\lambda)
  =\frac12\bm\lambda^{\T}\bm H\bm\lambda,
  \qquad
  R_a=\sigma_{0a}+\sum_b H_{ab}\lambda_b.
  \label{eq:coupled_hardening}
\end{equation}
The diagonal terms describe self-hardening and the off-diagonal terms describe latent or coupled hardening.

Kinematic hardening is represented by strain-like variables $\bm a_r$ and a storage potential $\psi^{\rm kin}(\bm a_1,\ldots,\bm a_{n_b})$. The backstresses are
\begin{equation}
  \bbeta_r=\frac{\partial\psi^{\rm kin}}{\partial\bm a_r},
  \qquad
  \bxi=\dev\bsig-\sum_r\bbeta_r.
  \label{eq:relative_stress}
\end{equation}
Linear kinematic hardening follows from a quadratic potential, while nonlinear
storage gives an energetic nonlinear translation. Dynamic recovery can be
included through a state-dependent tangent and its conjugate dissipative work
\citep{ArmstrongFrederick1966,Chaboche1986,Chaboche1989,Chaboche2008}.

\topic{Dynamic recovery and multiple backstresses}
Using stress-like backstresses $\balpha_i$, take
\begin{equation}
  \psi^{\rm kin}
  =\sum_{i=1}^{n_b}\frac{3}{4C_i}\balpha_i:\balpha_i,
  \qquad
  \bxi=\dev\bsig-\sum_i\balpha_i,
  \qquad
  q=\sqrt{\frac32\bxi:\bxi},
  \qquad
  \bN=\frac32\frac{\bxi}{q}.
  \label{eq:chaboche_kin_energy}
\end{equation}
For an accumulated plastic activity $p$, an Armstrong--Frederick-type tangent is
\begin{equation}
  \delta\bepsp=\bN\,\delta p,
  \qquad
  \delta\balpha_i
  =\left(\frac23C_i\bN-\gamma_i\balpha_i\right)\delta p.
  \label{eq:AF}
\end{equation}
The recovery term contributes the nonnegative directional work
\begin{equation}
  D_{\rm rec}
  =\sum_i\frac{3\gamma_i}{2C_i}\balpha_i:\balpha_i.
  \label{eq:ss_chaboche_recovery_cost}
\end{equation}
Pairing this term with the corresponding one-homogeneous recovery resistance
leaves the standard reduced force
\begin{equation}
  F=q-[\sigma_y+R(p)],
  \qquad
  R(p)=Q[1-\exp(-bp)]
  \label{eq:chaboche_yield}
\end{equation}
for a Voce isotropic hardening law. Under the optional time parameterization,
\Cref{eq:AF} becomes the usual Armstrong--Frederick/Chaboche evolution. The
backward-Euler update is recorded in Appendix~\ref{app:ss_discrete}.

\topic{Representative choices of \texorpdfstring{$W$}{W} and \texorpdfstring{$\psi^p$}{psi p}}
\label{subsec:ss_functional_choices}

Changing individual terms of the functional recovers standard hardening laws.
For a single activity, $R=W_{,\lambda}$; representative choices are
\begin{align}
  W_{\rm lin}
  &=\sigma_0\lambda+\frac12H\lambda^2,
  &R_{\rm lin}&=\sigma_0+H\lambda,
  \\
  W_{\rm Voce}
  &=\sigma_0\lambda
    +Q\left[\lambda+\frac{e^{-b\lambda}-1}{b}\right],
  &R_{\rm Voce}&=\sigma_0+Q(1-e^{-b\lambda}),
  \\
  W_{\rm pow}
  &=\sigma_0\lambda+\frac{K}{n+1}\lambda^{n+1},
  &R_{\rm pow}&=\sigma_0+K\lambda^n.
  \label{eq:ss_resistance_choices}
\end{align}
These choices recover linear, saturating, and power-law isotropic hardening
through the derivative $R=W_{,\lambda}$.
For several activities,
\begin{equation}
  W(\bm\lambda)
  =\sum_a\sigma_{0a}\lambda_a
   +\frac12\bm\lambda^{\T}\bm H\bm\lambda
\end{equation}
produces self-hardening from the diagonal of $\bm H$ and latent hardening from
its off-diagonal entries.

Kinematic forces follow from the plastic-state energy.  The choice
\begin{equation}
  \psi^p=\frac12\bepsp:\mathbb H:\bepsp
\end{equation}
produces the linear energetic backstress
$\bsig_p=\mathbb H:\bepsp$.  A spectrum of backstresses is obtained by
introducing separate strain-like variables and a quadratic form in those
variables; summing quadratic terms in the same $\bepsp$ would only change the
single effective modulus.  A nonlinear $\psi^p$ produces a nonlinear
energetic translation.  A state-dependent tangent relation may additionally generate
dynamic recovery, as shown later for Armstrong--Frederick evolution.
Finally,
\begin{equation}
  W(\lambda,\nabla\lambda)
  =W_0(\lambda)+\frac12c_\lambda|\nabla\lambda|^2
\end{equation}
produces the effective resistance
$R^{\rm eff}=W_{0,\lambda}-\nabla\!\cdot(c_\lambda\nabla\lambda)$ and its natural
micro-boundary condition.  The constitutive specialization changes, while the directional-stationarity
structure remains the same.

\topic{Independent activity variables in multisurface plasticity}

Suppose that several mechanisms share one scalar accumulated variable $\lambda$ and satisfy
\begin{equation}
  \delta\bm z_I
  =
  \left(\sum_a c_a\mathcal M_{Ia}\right)\delta\lambda.
\end{equation}
The directional variation then produces one combined directional force,
\begin{equation}
  F
  =
  \sum_I\mathcal X_I\mathbin{\bullet}
  \left(\sum_a c_a\mathcal M_{Ia}\right)-R(\lambda).
\end{equation}
A single nonnegative variation activates the combined mechanism; independent loading and unloading require separate activity variables.

By contrast, independent variables $\lambda_a$ generate independent inequalities \Cref{eq:admissibility_multi} and an active set
\begin{equation}
  \mathcal A=\{a\,|\,F_a=0\}.
\end{equation}
For an associated Koiter-type model, let each mechanism possess an admissible
unit set with support function $G_a$.  The directional variation first selects
$\bm N_a\in\partial_{\bsig}G_a$ and gives $F_a=G_a-R_a$.  At a smooth point,
$\bm N_a=F_{a,\bsig}$, so the admissible variation reduces to
\begin{equation}
  \delta\bepsp
  =
  \sum_{a\in\mathcal A}\delta\lambda_a
  \frac{\partial F_a}{\partial\bsig}.
  \label{eq:koiter}
\end{equation}
The selected support directions give the conic normality form.  For crystal plasticity, each slip system has its own accumulated
slip and the hardening matrix in \Cref{eq:coupled_hardening} becomes the
latent-hardening matrix. The formulation also accommodates nested cyclic surfaces and other coupled multisurface mechanisms.

\subsection{Multi-field and gradient variation}
\label{sec:ss_multifield}

Gradient plasticity has been formulated through plastic-strain gradients,
accumulated-strain gradients, defect energies, and microforce balances
\citep{Aifantis1984,FleckHutchinson1997,FleckHutchinson2001,Gurtin2000,
GurtinAnand2005,GurtinAnand2009,FleckWillis2009}. Recent work has further
examined saturating internal variables, elastic-gap-free decompositions, and
variational thermomechanical coupling
\citep{TeichtmeisterKeip2022,AbatourForest2024Saturating,
MukherjeeBanerjee2024ElasticGap}. In the present framework the bulk resistance
and natural micro-boundary term arise from one first variation, so
the regularization and its boundary condition remain linked.

The gradient contribution follows from an
explicit functional.  The derivation shows how coupled hardening terms, gradient
resistance, boundary conditions, and independent complementarity relations
originate.

Let $\bm z_I$, $I=1,\ldots,n_z$, be additional strain-like internal variables
and let $c_{ab}=c_{ba}$ be a positive-semidefinite gradient matrix.  Consider
\begin{align}
  \Pi_t[\bm u,\bepsp,\{\bm z_I\},\bm\lambda]
  ={}&\int_\Omega\Bigg[
    \phi(\beps-\bepsp)
    +\Psi(\bepsp,\bm z_1,\ldots,\bm z_{n_z})
    +\sum_{a=1}^{N}\sigma_{0a}\lambda_a
    +H(\bm\lambda)
    \notag\\
    &\hspace{29mm}
    +\frac12\sum_{a,b=1}^{N}c_{ab}
       \nabla\lambda_a\cdot\nabla\lambda_b
  \Bigg]\dd x-\ell_t(\bm u).
  \label{eq:ss_multifield_functional}
\end{align}
Define the energetic forces
\begin{equation}
  \bsig=\phi_{,\beps_e},
  \qquad
  \bxi=\bsig-\Psi_{,\bepsp},
  \qquad
  \mathcal X_I=-\Psi_{,\bm z_I},
  \qquad
  Q_a=H_{,\lambda_a}.
  \label{eq:ss_multifield_forces}
\end{equation}
The signs are selected so that $\bxi$ and $\mathcal X_I$ perform positive work
on their admissible internal-variable directions.  The first
variation is
\begin{align}
  \delta\Pi_t
  ={}&\delta\Pi_{\rm eq}
  -\int_\Omega\left[
      \bxi:\delta\bepsp
      +\sum_I\mathcal X_I\mathbin{\bullet}\delta\bm z_I
    \right]\dd x
  \notag\\
  &+\int_\Omega\sum_a(\sigma_{0a}+Q_a)\delta\lambda_a\,\dd x
  +\int_\Omega\sum_{a,b}c_{ab}\nabla\lambda_b\cdot
       \nabla\delta\lambda_a\,\dd x .
  \label{eq:ss_multifield_first_variation_weak}
\end{align}
This expression is then restricted to admissible plastic directions.

The admissible tangent cone is specified once by
\begin{equation}
  \delta\bepsp=\sum_a\bm N_a\delta\lambda_a,
  \qquad
  \delta\bm z_I=\sum_a\bm M_{Ia}\delta\lambda_a,
  \qquad
  \delta\lambda_a\ge0,
  \label{eq:ss_multifield_tangent}
\end{equation}
where $\bm N_a$ may be $G_{a,\bxi}$ and the other components may likewise be
derivatives of a scalar direction generator in generalized-force space.  After
substitution and integration by parts,
\begin{align}
  \delta\Pi_t
  ={}&\delta\Pi_{\rm eq}
  -\int_\Omega\sum_a F_a\delta\lambda_a\,\dd x
  +\int_{\partial\Omega}\sum_a t_a^\lambda\delta\lambda_a\,\dd s,
  \label{eq:ss_multifield_directional_variation}\\
  F_a={}&
  \bxi:\bm N_a
  +\sum_I\mathcal X_I\mathbin{\bullet}\bm M_{Ia}
  -R_a^{\rm eff},
  \label{eq:ss_multifield_obstruction}\\
  R_a^{\rm eff}={}&
  \sigma_{0a}+Q_a
  -\operatorname{div}\left(\sum_b c_{ab}\nabla\lambda_b\right),
  \label{eq:ss_multifield_effective_resistance}\\
  t_a^\lambda={}&
  \left(\sum_b c_{ab}\nabla\lambda_b\right)\cdot\bm n.
  \label{eq:ss_multifield_microtraction}
\end{align}
The boundary term generates either the natural micro-free condition
$t_a^\lambda=0$ or an essential micro-hard condition
$\delta\lambda_a=0$.  The integration by parts supplies both the bulk gradient resistance and the
corresponding micro-boundary term \citep{Aifantis1984,FleckHutchinson2001,GurtinAnand2005}.

Because the $\delta\lambda_a$ are independent and one-sided, the bulk
optimality conditions are
\begin{equation}
  F_a\le0,
  \qquad
  \delta\lambda_a\ge0,
  \qquad
  F_a\,\delta\lambda_a=0,
  \qquad a=1,\ldots,N.
  \label{eq:ss_multifield_KKT}
\end{equation}
Thus $N$ accumulated variables generate $N$ directional forces and $N$
plastic multipliers.  Cross terms in $H$ or $c_{ab}$ couple those mechanisms
while preserving their independent loading and unloading decisions.

\topic{Consistency matrix on an active set}

The multiplier equations follow by varying the active equalities.  To show the
structure, take quadratic elasticity, fixed local directions $\bm N_a$, no
additional $\bm z_I$, and local coupled hardening
$H=\frac12\bm\lambda^{\T}\bm H\bm\lambda$.  Then
\begin{equation}
  \delta\bsig
  =\mathbb C:\left(
    \delta\beps-\sum_b\bm N_b\delta\lambda_b
  \right)
\end{equation}
and, on an active set $\mathcal A$,
\begin{align}
  0=\delta F_a
  ={}&\bm N_a:\mathbb C:\delta\beps
  -\sum_{b\in\mathcal A}
  \left(
    \bm N_a:\mathbb C:\bm N_b+H_{ab}
  \right)\delta\lambda_b,
  \qquad a\in\mathcal A.
  \label{eq:ss_active_consistency_matrix}
\end{align}
Hence
\begin{equation}
  \sum_{b\in\mathcal A}\mathcal H_{ab}\delta\lambda_b
  =\bm N_a:\mathbb C:\delta\beps,
  \qquad
  \mathcal H_{ab}=\bm N_a:\mathbb C:\bm N_b+H_{ab}.
  \label{eq:ss_active_multiplier_system}
\end{equation}
For associated fixed directions and symmetric $\bm H$, the active matrix is
symmetric.  State-dependent or non-associated directions add the derivatives of
the directions and generally make the matrix nonsymmetric.  In either case the
system is the derivative of the active equalities $F_a=0$.

\topic{Single-field gradient plasticity}

For one activity and
\begin{equation}
  W(\lambda,\nabla\lambda)
  =\sigma_0\lambda+\frac12H\lambda^2
   +\frac12c\lvert\nabla\lambda\rvert^2,
  \qquad c\ge0,
  \label{eq:ss_single_gradient_W}
\end{equation}
\Cref{eq:ss_multifield_effective_resistance} reduces to
\begin{equation}
  R^{\rm eff}=\sigma_0+H\lambda-c\nabla^2\lambda.
  \label{eq:ss_single_gradient_R}
\end{equation}
The scalar functional then yields
\begin{equation}
  F_G=\bxi:G_{,\bxi}
      -\left(\sigma_0+H\lambda-c\nabla^2\lambda\right),
  \qquad
  c\nabla\lambda\cdot\bm n=0
  \quad\hbox{on a micro-free boundary}.
  \label{eq:ss_single_gradient_F}
\end{equation}
Variation of the gradient term produces both the Laplacian contribution and the
natural micro-boundary condition.

\subsection{Two-surface example with two accumulated variables}

This example is the smallest setting in which independent activities, a
coupled hardening matrix, and a simultaneous active set can be seen
explicitly \citep{Koiter1953,Iwan1967,SimoKennedyGovindjee1988}.

\label{sec:ss_two_surface_example}

A scalar tension--compression example makes the variational role of several
$\lambda$ variables explicit.  Let
\begin{align}
  \Pi[u,\varepsilon^p,\lambda_+,\lambda_-]
  =\int_\Omega\Bigg[&
    \frac12E(\varepsilon-\varepsilon^p)^2
    +\frac12K(\varepsilon^p)^2
    +\sigma_t\lambda_+
    +\sigma_c\lambda_-
    \notag\\
    &+\frac12H_{++}\lambda_+^2
    +H_{+-}\lambda_+\lambda_-
    +\frac12H_{--}\lambda_-^2
  \Bigg]\dd x-\ell(u).
  \label{eq:ss_two_surface_functional}
\end{align}
The energetic forces are
\begin{equation}
  \sigma=E(\varepsilon-\varepsilon^p),
  \qquad
  \sigma_p=K\varepsilon^p,
  \qquad
  \xi=\sigma-\sigma_p.
  \label{eq:ss_two_surface_forces}
\end{equation}
Choose the two one-sided tangent directions
\begin{equation}
  \delta\varepsilon^p=\delta\lambda_+-\delta\lambda_-,
  \qquad
  \delta\lambda_+\ge0,
  \quad
  \delta\lambda_-\ge0.
  \label{eq:ss_two_surface_tangent}
\end{equation}
The variation becomes
\begin{equation}
  \delta\Pi
  =\delta\Pi_{\rm eq}
   -\int_\Omega
   \left(F_+\delta\lambda_++F_-\delta\lambda_-\right)\dd x,
  \label{eq:ss_two_surface_variation}
\end{equation}
with two directional forces,
\begin{align}
  F_+
  &=\xi-\left(\sigma_t+H_{++}\lambda_++H_{+-}\lambda_-\right),
  \label{eq:ss_tension_obstruction}\\
  F_-
  &=-\xi-\left(\sigma_c+H_{--}\lambda_-+H_{+-}\lambda_+\right).
  \label{eq:ss_compression_obstruction}
\end{align}
One-sided stability gives
\begin{equation}
  F_\pm\le0,
  \qquad
  \delta\lambda_\pm\ge0,
  \qquad
  F_\pm\,\delta\lambda_\pm=0.
  \label{eq:ss_two_surface_KKT}
\end{equation}
The tensile and compressive mechanisms can now load and unload independently,
while $H_{+-}$ produces latent coupling.  Replacing $(\lambda_+,\lambda_-)$ by
a single scalar would collapse \Cref{eq:ss_tension_obstruction,eq:ss_compression_obstruction}
into one weighted directional force and eliminate that independent active-set
structure.  This elementary calculation is the direct prototype of Koiter
multi-surface plasticity, crystal slip, and coupled shear--cap mechanisms.

\subsection{Tensorial example: mixed-hardening \texorpdfstring{$J_2$}{J2} plasticity}

The tensorial example recovers the classical distortion-energy family and the
Prandtl--Reuss flow structure before adding mixed hardening
\citep{Huber1904,Hencky1924,Prandtl1924,Reuss1930,Hill1950}.

\label{sec:ss_example}

This example starts with a continuous functional and carries out its first
variation to derive the constitutive relations. It recovers the classical
Huber--von Mises--Hencky surface and Prandtl--Reuss direction
\citep{Huber1904,vonMises1913,Hencky1924,Prandtl1924,Reuss1930}, while the
isotropic resistance and kinematic translation remain identifiable as separate
derivatives of $W$ and $\psi^p$. The time-discrete return map is presented
separately from the continuum theory.

\topic{Functional and constitutive forces}

Let
\begin{equation}
  \Pi_t[\bm u,\bepsp,p]
  =\int_\Omega
  \left[
    \frac12(\beps-\bepsp):\mathbb C:(\beps-\bepsp)
    +\frac13C_k\bepsp:\bepsp
    +\sigma_y p+\frac12H_i p^2
  \right]\dd x-\ell_t(\bm u).
  \label{eq:ss_example_energy}
\end{equation}
The scalar functional contains linear elasticity, Prager kinematic hardening,
linear isotropic hardening, and initial resistance.  Differentiation
gives
\begin{equation}
  \bsig=\mathbb C:(\beps-\bepsp),
  \qquad
  \bbeta=\frac23C_k\bepsp,
  \qquad
  R=\sigma_y+H_i p,
  \label{eq:ss_example_forces}
\end{equation}
and the relative deviatoric force is
\begin{equation}
  \bxi=\dev\bsig-\bbeta,
  \qquad
  q(\bxi)=\sqrt{\frac32\bxi:\bxi}.
  \label{eq:ss_example_relative_force}
\end{equation}
The plastic metric is specified by the normalized deviatoric unit set
\begin{equation}
  \mathcal U_{J_2}
  =\left\{\bm N:\tr\bm N=0,
  \ \sqrt{\frac23\bm N:\bm N}\le1\right\}.
  \label{eq:ss_example_j2_unit_set}
\end{equation}
Its support function is the von Mises gauge,
\begin{equation}
  G(\bxi)
  =\sup_{\bm N\in\mathcal U_{J_2}}\bxi:\bm N
  =q(\bxi).
  \label{eq:ss_example_tangent_generator}
\end{equation}
For $q>0$, minimization of the restricted first variation selects the unique
support direction
\begin{equation}
  \bN
  =\operatorname*{arg\,max}_{\bm M\in\mathcal U_{J_2}}\bxi:\bm M
  =G_{,\bxi}=q_{,\bxi}
  =\frac32\frac{\bxi}{q},
  \qquad
  \bxi:\bN=q,
  \qquad
  \bN:\bN=\frac32.
  \label{eq:ss_example_normal}
\end{equation}
The selected one-sided plastic variation is
\begin{equation}
  \delta\bepsp=\bN\,\delta p,
  \qquad \delta p\ge0.
  \label{eq:ss_example_direction_variation}
\end{equation}
Thus the $J_2$ metric is constitutive input, while the strength measure and the
associated direction are the support value and support direction generated by
the same directional minimization.

\topic{First variation and continuum relations}

The first variation is
\begin{align}
  \delta\Pi_t
  ={}&\int_\Omega
  \left[
    \bsig:\delta\beps
    -\bsig:\delta\bepsp
    +\bbeta:\delta\bepsp
    +(\sigma_y+H_i p)\delta p
  \right]\dd x-\delta\ell_t
  \notag\\
  ={}&\delta\Pi_{\rm eq}
  -\int_\Omega
  \left[
    (\dev\bsig-\bbeta):\bN
    -\sigma_y-H_i p
  \right]\delta p\,\dd x
  \notag\\
  ={}&\delta\Pi_{\rm eq}
  -\int_\Omega
  \left[q-\sigma_y-H_i p\right]\delta p\,\dd x.
  \label{eq:ss_example_first_variation}
\end{align}
Therefore the directional force for this functional is
\begin{equation}
  \boxed{F=q-\sigma_y-H_i p.}
  \label{eq:ss_example_activation_force}
\end{equation}
Directional stationarity on the one-sided activity cone gives
\begin{equation}
  F\le0,
  \qquad
  \delta p\ge0,
  \qquad
  F\,\delta p=0,
  \label{eq:ss_example_kkt}
\end{equation}
and the admissible tangent gives
\begin{equation}
  \delta\bepsp=\bN\,\delta p,
  \qquad
  \delta\bbeta=\frac23C_k\bN\,\delta p,
  \qquad
  \delta R=H_i\,\delta p.
  \label{eq:ss_example_continuous_evolution}
\end{equation}
The elastic law, backstress, isotropic resistance, active boundary,
complementarity conditions, hardening variations, and consistency relation are
therefore obtained before time discretization.  The von Mises gauge selects $\bN$ as its support direction, and the relation
$\bN=F_{,\bxi}$ is verified below.

The associated character is checked against the boundary obtained from the variation:
\begin{equation}
  F_{,\bxi}=q_{,\bxi}=\bN.
  \label{eq:ss_example_self_similarity}
\end{equation}
The degree-one identity $\bxi:\bN=q$ explains why one scalar gauge can label both the tangent and the resulting boundary in this example.  Normality is therefore a consequence of the directional criterion $\bM\parallel F_{,\bxi}$; equality of scalar labels is not a definition of association.

\topic{Loading, unloading, and active consistency}

If \(F<0\), complementarity gives \(\delta p=0\), so an admissible stress
variation is elastic.  If \(\delta p>0\), the functional is stationary in the
active direction and \(F=0\).  Variation of this active equality gives
\begin{equation}
  \delta F
  =\bN:\delta\bxi-H_i\,\delta p=0.
  \label{eq:ss_example_consistency_rate}
\end{equation}
For isotropic elasticity,
\begin{equation}
  \delta\bxi
  =2G_s\dev\delta\beps
   -\left(2G_s+\frac23C_k\right)\bN\,\delta p,
\end{equation}
where \(G_s\) is the elastic shear modulus.  Using
\(\bN:\bN=3/2\) yields
\begin{equation}
  \delta p
  =\frac{\bN:2G_s\dev\delta\beps}
  {3G_s+C_k+H_i}
  \quad\hbox{on an active loading variation},
  \label{eq:ss_example_multiplier_rate}
\end{equation}
with \(\delta p=0\) whenever the numerator would violate the one-sided
condition.

\topic{Time-discrete update}
The backward-Euler radial return and its closed-form multiplier are given in Appendix~\ref{app:ss_discrete}.

\topic{A non-associated variation}

To obtain a nontrivial volumetric direction, introduce the full relative force
\begin{equation}
  \boldsymbol\eta=\bsig-\bbeta,
  \qquad
  q(\boldsymbol\eta)
  =\sqrt{\frac32\dev\boldsymbol\eta:\dev\boldsymbol\eta}.
  \label{eq:ss_example_full_relative_force}
\end{equation}
The activity variable in this illustration is a path coordinate distinct from the conventional equivalent plastic strain.  Choose the tangent generator
\begin{equation}
  G_B(\boldsymbol\eta)
  =q(\boldsymbol\eta)
   +\frac{a}{2\sigma_*}(\tr\boldsymbol\eta)^2,
  \qquad a\ge0,\quad\sigma_*>0,
  \label{eq:ss_example_nonassoc_direction}
\end{equation}
and restrict the plastic variation by
$\delta\bepsp=G_{B,\boldsymbol\eta}\,\delta p$.  The resistance remains
$\sigma_y+H_i p$.  Since
\begin{equation}
  G_{B,\boldsymbol\eta}
  =\frac32\frac{\dev\boldsymbol\eta}{q}
   +\frac{a}{\sigma_*}(\tr\boldsymbol\eta)\bm I,
\end{equation}
the directional variation gives
\begin{equation}
  \boxed{
  F_B
  =q+\frac{a}{\sigma_*}(\tr\boldsymbol\eta)^2
   -\sigma_y-H_i p.}
  \label{eq:ss_example_nonassoc_yield}
\end{equation}
The normal to the resulting boundary contains
$2(a/\sigma_*)(\tr\boldsymbol\eta)\bm I$, whereas the selected tangent contains
only $(a/\sigma_*)(\tr\boldsymbol\eta)\bm I$.  The two directions are therefore
not parallel except on special stress states.  This example also shows why the
full relative force is required: the trace term would vanish identically if it
were written in terms of the deviatoric force $\bxi$ used in the preceding
von Mises specialization.

\topic{Summary of the tensorial example}

The constitutive inputs are the scalar functional, the material parameters,
and the admissible tangent generator. Their first and directional variations
give stress, backstress, isotropic resistance, the directional force,
complementarity conditions, hardening variations, and the consistency equation.
Equivalently, a material is specified by the pair
\begin{equation}
  \mathfrak M=(\Pi_t,\mathcal K),
  \label{eq:ss_constitutive_pair}
\end{equation}
where the functional determines stresses and resistance derivatives and
\(\mathcal K\) specifies the one-sided admissible directions. Time-discrete
integration is collected in Appendix~\ref{app:ss_discrete}.

\section{Analytical examples}
\label{sec:ss_analytical_solutions}

The closed-form examples below emphasize different aspects of the formulation. Multi-threshold torsion produces several radial active regions, the non-uniform bar develops a spatial backstress field, the Hill-type annulus provides a classical pressure-sensitive limit state, and cavity expansion separates pressure-dependent strength from plastic dilatancy within a two-dimensional boundary-value problem.

\subsection{Multi-threshold torsion of a circular annulus}
\label{sec:ss_polar_multilambda_solution}

\begin{figure}[t]
  \centering
  \includegraphics[width=0.98\textwidth]{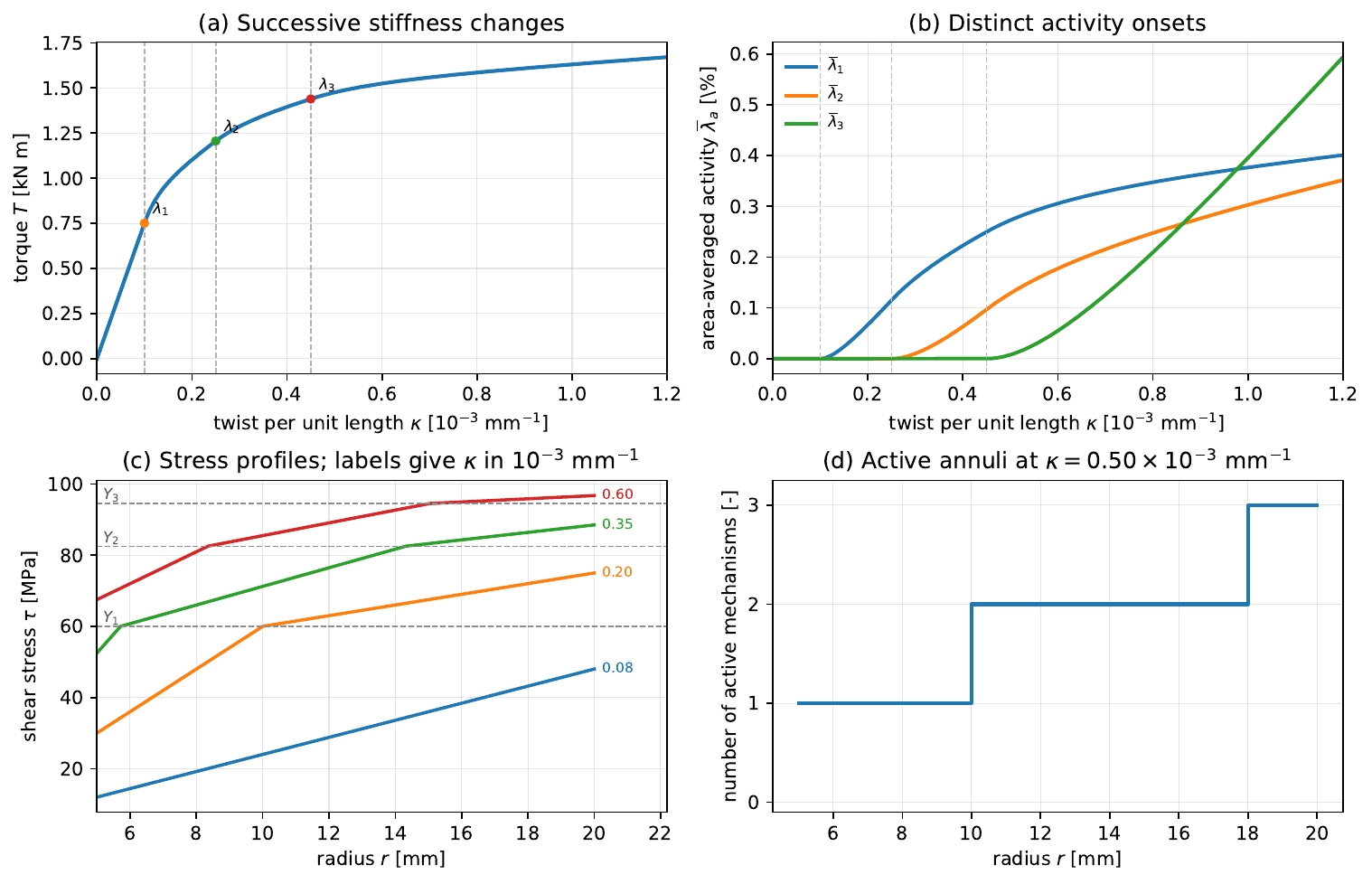}
  \caption{Multi-threshold torsion of a circular annulus.  (a) Torque--twist
  response with the three outer-surface activation points marked; (b)
  area-averaged activities, showing distinct onsets; (c) radial shear-stress
  profiles and the three thresholds; and (d) nested active annuli at
  \(\kappa=0.5\times10^{-3}~\mathrm{mm}^{-1}\).  The parameters are
  \(R_i=5~\mathrm{mm}\), \(R_o=20~\mathrm{mm}\),
  \(G=30~\mathrm{GPa}\), \(Y=(60,82.5,94.5)~\mathrm{MPa}\), and
  \(H=(10,5,1)~\mathrm{GPa}\).}
  \label{fig:ss_analytical_torsion}
\end{figure}

Consider a long circular annulus with inner radius $R_i$ and outer radius
$R_o$ under monotone positive torsion.  The twist per unit length $\kappa$ is
the generalized coordinate and $T$ is its conjugate torque; under displacement
control the same relation gives $T$ as the reaction.  The engineering shear
strain is
\begin{equation}
  \gamma(r)=\kappa r,
  \qquad R_i\le r\le R_o .
  \label{eq:ss_polar_gamma}
\end{equation}
Introduce activities $\lambda_a(r)\ge0$ in the same shear direction, with
thresholds $Y_a$ and hardening moduli $H_a>0$.  The functional per unit length is
\begin{equation}
  \Pi_\kappa
  =2\pi\int_{R_i}^{R_o}
  \left[
  \frac12 G\left(\kappa r-\sum_{a=1}^{N}\lambda_a\right)^2
  +\sum_{a=1}^{N}\left(Y_a\lambda_a+\frac12H_a\lambda_a^2\right)
  \right] r\,\dd r
  -T\kappa .
  \label{eq:ss_polar_functional}
\end{equation}
The shear stress is
\begin{equation}
  \tau(r)=G\left(\kappa r-\sum_{a=1}^{N}\lambda_a(r)\right).
  \label{eq:ss_polar_tau}
\end{equation}
The first variation gives
\begin{equation}
  T=2\pi\int_{R_i}^{R_o}r^2\tau(r)\,\dd r,
  \qquad
  F_a(r)=\tau(r)-Y_a-H_a\lambda_a(r),
  \label{eq:ss_polar_directional_force}
\end{equation}
and, for monotone loading from the virgin state, the local optimality conditions
\begin{equation}
  F_a(r)\le0,
  \qquad
  \lambda_a(r)\ge0,
  \qquad
  F_a(r)\lambda_a(r)=0 .
  \label{eq:ss_polar_kkt}
\end{equation}
On an interval with active set $\mathcal A$, the solution is
\begin{equation}
  \lambda_a(r)=\frac{\tau(r)-Y_a}{H_a}\quad(a\in\mathcal A),
  \qquad
  \tau_{\mathcal A}(r)
  =\frac{G\kappa r+G\displaystyle\sum_{a\in\mathcal A}Y_a/H_a}
  {1+G\displaystyle\sum_{a\in\mathcal A}1/H_a} .
  \label{eq:ss_polar_active_stress}
\end{equation}
For $G>0$ and $H_a>0$, the local density is strictly convex in the activities,
so this active-set solution is unique.  For $\kappa>0$ and ordered thresholds $0<Y_1<\cdots<Y_N$, activation radii follow from
$\tau_{\mathcal A_{j-1}}(r_j)=Y_j$:
\begin{equation}
  r_j(\kappa)
  =\frac{Y_j(1+GS_{j-1})-GB_{j-1}}{G\kappa},
  \qquad
  S_j=\sum_{a=1}^{j}\frac1{H_a},
  \qquad
  B_j=\sum_{a=1}^{j}\frac{Y_a}{H_a},
  \qquad S_0=B_0=0 .
  \label{eq:ss_polar_activation_radius}
\end{equation}
When the resulting activation radii are ordered, set $\rho_0=R_i$,
$\rho_{N+1}=R_o$, and clip $\rho_j=r_j$ to $[R_i,R_o]$.  If the radii are
not ordered, the intervals are obtained by direct evaluation of the local
active set.  For the ordered case, the torque is
\begin{equation}
  T(\kappa)
  =2\pi\sum_{j=0}^{N}
  \left[
  \frac{a_j\kappa}{4}\left(\rho_{j+1}^4-\rho_j^4\right)
  +\frac{b_j}{3}\left(\rho_{j+1}^3-\rho_j^3\right)
  \right],
  \quad
  a_j=\frac{G}{1+GS_j},\quad
  b_j=\frac{GB_j}{1+GS_j} .
  \label{eq:ss_polar_closed_torque}
\end{equation}
The fields in \Cref{fig:ss_analytical_torsion} show the successive activation of the three mechanisms.  The annulus has
\(R_i=5~\mathrm{mm}\), \(R_o=20~\mathrm{mm}\), and
\(G=30~\mathrm{GPa}\).  The thresholds are
\(Y=(60,82.5,94.5)~\mathrm{MPa}\), and the hardening moduli are
\(H=(10,5,1)~\mathrm{GPa}\).  These parameters separate the three onset
points while retaining nested radial active regions.  The area-averaged
activities plotted against twist are
\begin{equation}
  \overline{\lambda}_a(\kappa)
  =\frac{2}{R_o^2-R_i^2}
  \int_{R_i}^{R_o}\lambda_a(r,\kappa)\,r\,\dd r.
  \label{eq:ss_polar_average_activity}
\end{equation}
The torque and averaged activities are shown for
\(0\le\kappa\le1.2\times10^{-3}~\mathrm{mm}^{-1}\); the radial active
set is shown at \(\kappa=0.5\times10^{-3}~\mathrm{mm}^{-1}\).

\subsection{Non-uniform bar with kinematic hardening}
\label{sec:ss_tapered_bar_kinematic_solution}

\begin{figure}[t]
  \centering
  \includegraphics[width=0.98\textwidth]{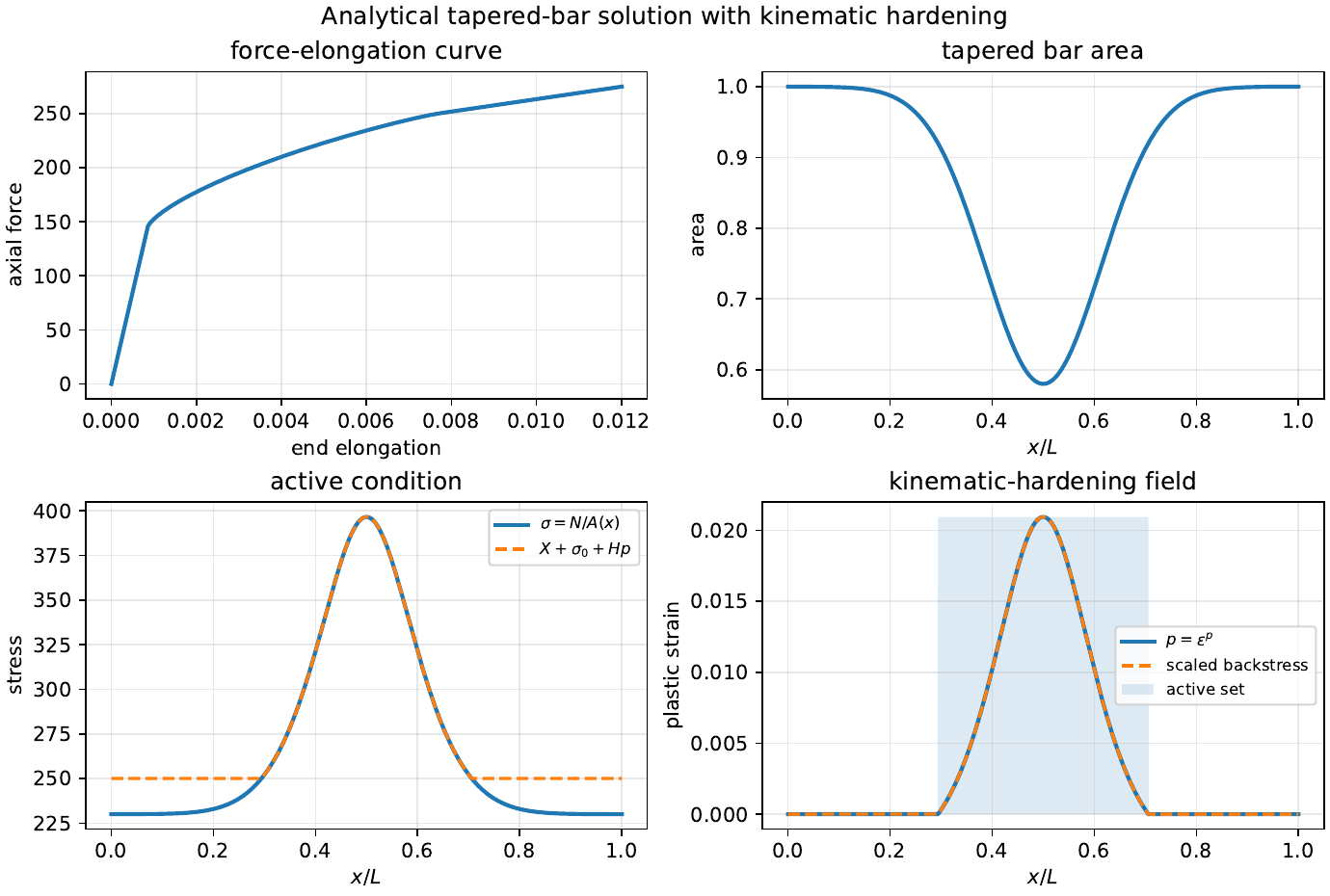}
  \caption{Non-uniform bar with linear kinematic hardening for
  \(L=1\), \(A(x)=1-0.42\exp[-((x-0.5)/0.16)^2]\),
  \(E=2.0\times10^5\), \(\sigma_0=250\), \(C=6000\), and
  \(H=1000\).  The field plots use \(N=230\); the active set is selected by
  \(A(x)<N/\sigma_0\).}
  \label{fig:ss_analytical_tapered_bar}
\end{figure}

Consider a bar of length $L$ and area $A(x)>0$, with $u(0)=0$ and an
axial force $N>0$ applied at $x=L$.  On a monotone tensile branch the plastic strain and accumulated activity
coincide, $p=\varepsilon^p\ge0$.  The functional is
\begin{equation}
  \Pi_N[u,p]
  =\int_0^L A(x)
  \left[
  \frac12 E(u'-p)^2
  +\sigma_0p
  +\frac12Hp^2
  +\frac12Cp^2
  \right]\dd x-Nu(L),
  \label{eq:ss_tapered_bar_functional}
\end{equation}
where $H$ is isotropic hardening and $C$ is linear kinematic hardening.  The
variation gives
\begin{equation}
  (A\sigma)'=0,
  \qquad
  \sigma=E(u'-p),
  \qquad
  X=Cp,
  \qquad
  R=\sigma_0+Hp,
  \label{eq:ss_tapered_bar_forces}
\end{equation}
and
\begin{equation}
  F(x)=\sigma(x)-X(x)-R(x)
       =\sigma(x)-\sigma_0-(C+H)p(x).
  \label{eq:ss_tapered_bar_force}
\end{equation}
Since $A(x)\sigma(x)=N$, the field is explicit:
\begin{equation}
  \sigma_N(x)=\frac{N}{A(x)},
  \qquad
  p_N(x)=\left\langle\frac{\sigma_N(x)-\sigma_0}{C+H}\right\rangle_+,
  \qquad
  X_N(x)=Cp_N(x).
  \label{eq:ss_tapered_bar_solution}
\end{equation}
The end elongation is
\begin{equation}
  U(N)=\int_0^L\left(\frac{\sigma_N(x)}{E}+p_N(x)\right)\dd x .
  \label{eq:ss_tapered_bar_elongation}
\end{equation}
The active region is selected by $A(x)<N/\sigma_0$.  On this monotone branch,
$C$ and $H$ enter the plastic strain through their sum, although the backstress
field $X=Cp$ remains separately identifiable; reverse loading is required to
distinguish their effects on the subsequent response.  \Cref{fig:ss_analytical_tapered_bar}
shows the area variation, the force--elongation curve, and the stress,
backstress, and plastic-strain fields.  The plotting data are
\(L=1\),
\(A(x)=1-0.42\exp[-((x-0.5)/0.16)^2]\),
\(E=2.0\times10^5\), \(\sigma_0=250\), \(C=6000\), and
\(H=1000\).  The field plots use \(N=230\); the force--elongation curve
uses \(0\le N\le275\).

\subsection{Hill-type annulus and a non-associated tangent}
\label{sec:ss_hill_nonassoc_solution}

\begin{figure}[t]
  \centering
  \includegraphics[width=0.98\textwidth]{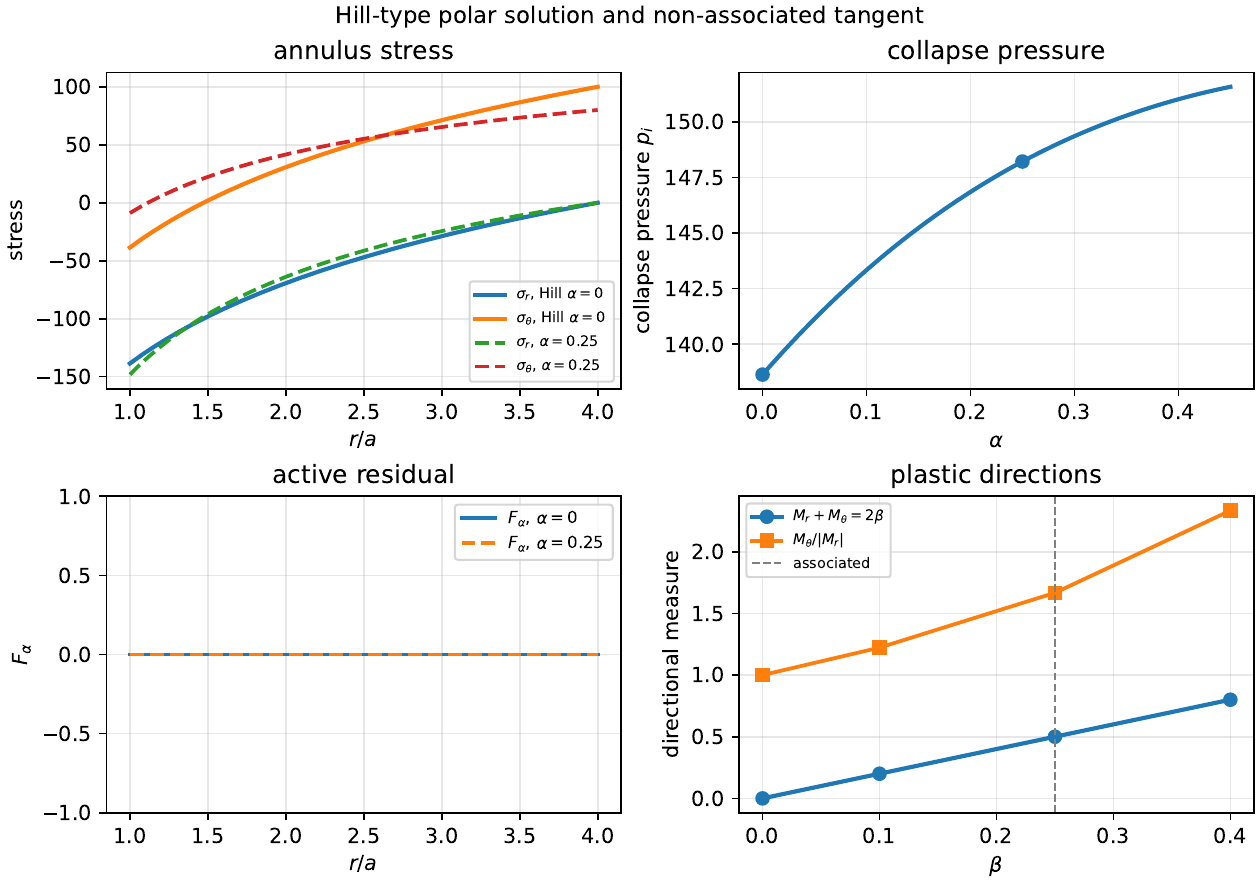}
  \caption{Hill-type annulus and non-associated tangent for
  \(a=1\), \(b=4\), \(k=100\), and \(p_o=0\).  The stress fields compare
  the Hill limit \(\alpha=0\) with \(\alpha=0.25\); the directional panel uses
  \(\beta=0,0.10,0.25,0.40\).}
  \label{fig:ss_analytical_hill}
\end{figure}

Let a thick annulus with inner radius $a$ and outer radius $b$ be loaded by
internal pressure $p_i$ and external pressure $p_o$.  In a fully plastic annulus
the polar equilibrium equation is
\begin{equation}
  \frac{\dd\sigma_r}{\dd r}+\frac{\sigma_r-\sigma_\theta}{r}=0 .
  \label{eq:ss_hill_equilibrium}
\end{equation}
Introduce
\begin{equation}
  d=\sigma_\theta-\sigma_r,
  \qquad
  s=\sigma_\theta+\sigma_r,
  \label{eq:ss_hill_ds}
\end{equation}
and consider the local generating density
$\phi(\beps-\bepsp)+k\lambda$ together with the admissible polar tangent
$\delta\bepsp=\bM_\alpha\,\delta\lambda$, where
\begin{equation}
  \bM_\alpha=(\alpha-1)\bm e_r\otimes\bm e_r
  +(\alpha+1)\bm e_\theta\otimes\bm e_\theta.
  \label{eq:ss_hill_associated_tangent}
\end{equation}
The directional first variation gives
\begin{equation}
  F_\alpha
  =\bsig:\bM_\alpha-k
  =d+\alpha s-k.
  \label{eq:ss_hill_boundary}
\end{equation}
In a fully plastic annulus $F_\alpha=0$.  The resulting limit solution gives
the statically admissible stress field and the corresponding pressure relation.
Elastic--plastic interfaces belong to the partially yielded extension of this
problem.  The closed form below assumes $1+\alpha\ne0$; the pressure-sensitive
cases plotted here have $\alpha>0$.
For $\alpha=0$ this reduces to the Hill constant stress-difference solution
\citep{Hill1950}.  For $\alpha\ne0$,
\begin{equation}
  \sigma_r(r)=\frac{k}{2\alpha}+C_0 r^{-m},
  \qquad
  \sigma_\theta(r)=\frac{k+(1-\alpha)\sigma_r(r)}{1+\alpha},
  \qquad
  m=\frac{2\alpha}{1+\alpha},
  \label{eq:ss_hill_alpha_solution}
\end{equation}
where $C_0=(-p_o-k/(2\alpha))b^m$, and
\begin{equation}
  p_i
  =-\frac{k}{2\alpha}
   +\left(p_o+\frac{k}{2\alpha}\right)\left(\frac{b}{a}\right)^m .
  \label{eq:ss_hill_alpha_pressure}
\end{equation}
The regular limit as $\alpha\to0$ is
\begin{equation}
  \sigma_r(r)=k\ln\frac{r}{b}-p_o,
  \qquad
  \sigma_\theta(r)=\sigma_r(r)+k,
  \qquad
  p_i=p_o+k\ln\frac{b}{a} .
  \label{eq:ss_hill_limit}
\end{equation}
A tangent defined by $G_\beta=d+\beta s$ has
\begin{equation}
  M_r=\beta-1,
  \qquad
  M_\theta=\beta+1 .
  \label{eq:ss_hill_tangent}
\end{equation}
The case $\beta=\alpha$ is associated.  If $\beta\ne\alpha$, the same active
boundary can be represented on a region where $d+\beta s>0$ by
\begin{equation}
  \widehat{\bM}_\beta
  =\frac{d+\alpha s}{d+\beta s}\bM_\beta,
  \qquad
  \bsig:\widehat{\bM}_\beta-k=d+\alpha s-k .
  \label{eq:ss_hill_power_normalized_tangent}
\end{equation}
This is a local representation of a non-associated tangent; it loses regularity
where the denominator approaches zero.  The unscaled plastic direction is
\begin{equation}
  \delta\varepsilon_r^p=(\beta-1)\delta\lambda,
  \qquad
  \delta\varepsilon_\theta^p=(\beta+1)\delta\lambda,
  \qquad
  \delta\varepsilon_r^p+\delta\varepsilon_\theta^p=2\beta\delta\lambda .
  \label{eq:ss_hill_flow_ratio}
\end{equation}
Thus \(\alpha\) controls the pressure-dependent boundary and \(\beta\) controls
the plastic dilatancy.  \Cref{fig:ss_analytical_hill} displays the stress fields
and directional measures.  The figure uses \(a=1\), \(b=4\), \(k=100\),
\(p_o=0\), the Hill limit \(\alpha=0\), a pressure-sensitive case
\(\alpha=0.25\), and the tangent parameters
\(\beta=0,0.10,0.25,0.40\).

\subsection{Elastic--plastic cavity expansion with independent strength and dilatancy}
\label{sec:ss_cavity_expansion_solution}

Cylindrical cavity expansion is a classical analytical test for pressure-sensitive plasticity and non-associated flow \citep{CarterBookerYeung1986,YuHoulsby1991,YuCarter2002,ChenWang2024Cavity}.  The following small-strain solution is formulated directly in the two-dimensional polar setting of the present theory.  It adds an elastic exterior, a moving elastic--plastic interface, and a displacement field to the fully plastic annulus of the preceding subsection.

Consider an infinite plane-strain medium containing a circular cavity of radius \(a\).  An internal pressure \(p\) produces a plastic annulus \(a\le r\le c\) and an elastic exterior \(r\ge c\).  Remote stress is zero and tension is positive.  Retain
\begin{equation}
  d=\sigma_\theta-\sigma_r,
  \qquad
  s=\sigma_\theta+\sigma_r,
  \qquad
  F_\alpha=d+\alpha s-k,
  \qquad 0<\alpha<1.
  \label{eq:ss_cavity_boundary}
\end{equation}
The positive normalization in \Cref{eq:ss_hill_power_normalized_tangent} permits the same boundary to be combined with a tangent parameter \(\beta\).  Absorbing the positive normalization into the activity coordinate \(\bar\lambda\) gives
\begin{equation}
  \delta\varepsilon_r^p=(\beta-1)\,\delta\bar\lambda,
  \qquad
  \delta\varepsilon_\theta^p=(\beta+1)\,\delta\bar\lambda,
  \qquad 0\le\beta<1.
  \label{eq:ss_cavity_flow}
\end{equation}
The choice \(\beta=\alpha\) is associated, whereas \(\beta\ne\alpha\) is non-associated.  The in-plane plastic dilatation is
\begin{equation}
  \delta\varepsilon_v^p
  =\delta\varepsilon_r^p+\delta\varepsilon_\theta^p
  =2\beta\,\delta\bar\lambda.
  \label{eq:ss_cavity_dilatancy}
\end{equation}
Thus \(\alpha\) controls the active strength boundary and \(\beta\) controls dilatancy.

In the elastic exterior, axisymmetric equilibrium and decay at infinity give
\begin{equation}
  \sigma_r^e=-\frac{k}{2}\left(\frac{c}{r}\right)^2,
  \qquad
  \sigma_\theta^e=\frac{k}{2}\left(\frac{c}{r}\right)^2,
  \qquad
  u^e(r)=\frac{(1+\nu)k c^2}{2Er}.
  \label{eq:ss_cavity_elastic_fields}
\end{equation}
At \(r=c\), \(F_\alpha=0\), so the elastic field supplies the interface data
\(\sigma_r(c)=-k/2\), \(\sigma_\theta(c)=k/2\), and
\(u(c)=(1+\nu)kc/(2E)\).

In the plastic annulus, radial equilibrium together with \(F_\alpha=0\) gives a closed stress field.  With
\begin{equation}
  x=\frac{r}{c},
  \qquad
  m=\frac{2\alpha}{1+\alpha},
  \label{eq:ss_cavity_xm}
\end{equation}
that field is
\begin{equation}
  \sigma_r^p
  =\frac{k}{2\alpha}\left[1-(1+\alpha)x^{-m}\right],
  \qquad
  \sigma_\theta^p
  =\frac{k}{2\alpha}\left[1-(1-\alpha)x^{-m}\right].
  \label{eq:ss_cavity_plastic_stress}
\end{equation}
The cavity pressure and plastic radius are therefore related by
\begin{equation}
  \frac{p}{k}
  =\frac{(1+\alpha)(c/a)^m-1}{2\alpha}.
  \label{eq:ss_cavity_pressure_radius}
\end{equation}
The regular limit \(\alpha\to0\) is
\(p/k=1/2+\ln(c/a)\), which recovers the constant stress-difference form.

The displacement distinguishes associated and non-associated response.  In the present plane-strain reduction,
\begin{equation}
  \varepsilon_r=u_{,r},
  \qquad
  \varepsilon_\theta=\frac{u}{r},
  \qquad
  \varepsilon_r^e=\frac{1+\nu}{E}\left[(1-\nu)\sigma_r-\nu\sigma_\theta\right],
  \qquad
  \varepsilon_\theta^e=\frac{1+\nu}{E}\left[(1-\nu)\sigma_\theta-\nu\sigma_r\right].
  \label{eq:ss_cavity_plane_strain_elasticity}
\end{equation}
Eliminating \(\bar\lambda\) from \Cref{eq:ss_cavity_flow} gives
\begin{equation}
  u_{,r}-n\frac{u}{r}
  =\varepsilon_r^e-n\varepsilon_\theta^e,
  \qquad
  n=\frac{\beta-1}{\beta+1}.
  \label{eq:ss_cavity_displacement_ode}
\end{equation}
Set
\begin{equation}
  C_0=\frac{(1+\nu)k}{2\alpha E},
  \qquad
  D=1-n-m
  =\frac{2(1-\alpha\beta)}{(1+\alpha)(1+\beta)},
  \label{eq:ss_cavity_auxiliary}
\end{equation}
and introduce the two dimensionless coefficients
\begin{equation}
  A_{\alpha\beta}
  =\frac{2\alpha(1-\nu)(1+\beta)}{1-\alpha\beta},
  \qquad
  B_{\alpha\beta}
  =\frac{(1+\alpha)(1+\alpha\beta-2\nu)}{1-\alpha\beta}.
  \label{eq:ss_cavity_amplitude}
\end{equation}
For \(0\le\alpha,\beta<1\), the denominator \(1-\alpha\beta\) is positive.  Continuity at \(r=c\) then gives
\begin{equation}
  \frac{u^p(r)}{c}
  =C_0\left[
    A_{\alpha\beta}x^n
    +(1-2\nu)x
    -B_{\alpha\beta}x^{1-m}
  \right].
  \label{eq:ss_cavity_displacement}
\end{equation}
The accumulated activity reduces to the compact expression
\begin{equation}
  \bar\lambda(r)
  =\frac{k(1-\nu^2)}{E(1-\alpha\beta)}
  \left(x^{n-1}-x^{-m}\right).
  \label{eq:ss_cavity_activity}
\end{equation}
Because \(n-1=-2/(1+\beta)\) and \(\alpha\beta<1\), \Cref{eq:ss_cavity_activity} is nonnegative for \(a\le r\le c\).  Writing \(q=c/a\), the cavity-wall displacement follows explicitly as
\begin{equation}
  \frac{u(a)}{a}
  =C_0\left[
    A_{\alpha\beta}q^{1-n}
    +(1-2\nu)
    -B_{\alpha\beta}q^m
  \right].
  \label{eq:ss_cavity_wall_displacement}
\end{equation}
Equations~\eqref{eq:ss_cavity_displacement}--\eqref{eq:ss_cavity_wall_displacement} satisfy
\(u^p(c)=u^e(c)\) and \(\bar\lambda(c)=0\).  The stress field and the pressure--plastic-radius relation depend on \(\alpha\), whereas the displacement, activity, and in-plane plastic dilatation depend on \(\beta\).  The solution therefore separates strength and flow kinematics within one boundary-value problem.

\begin{figure}[t]
  \centering
  \includegraphics[width=0.98\textwidth]{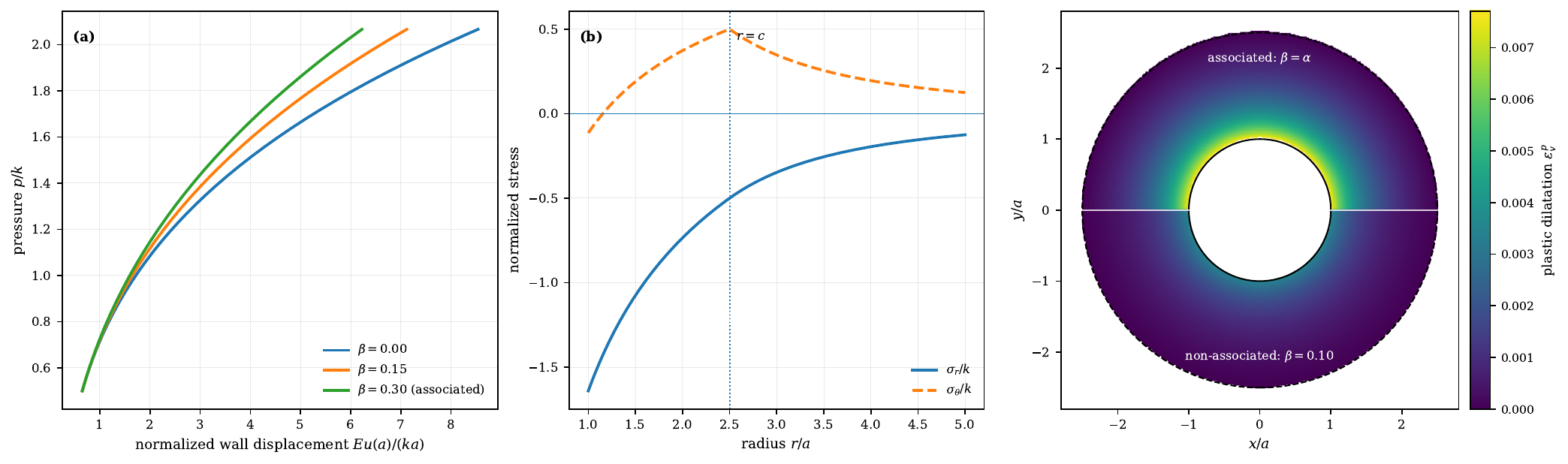}
  \caption{Elastic--plastic cavity expansion for \(E/k=200\), \(\nu=0.30\), and \(\alpha=0.30\): (a) pressure versus normalized cavity-wall displacement for three dilatancy parameters; (b) radial and hoop stresses at \(c/a=2.5\); and (c) in-plane plastic dilatation, with associated flow in the upper half and \(\beta=0.10\) in the lower half.}
  \label{fig:ss_analytical_cavity}
\end{figure}

\Cref{fig:ss_analytical_cavity} uses \(E/k=200\), \(\nu=0.30\), and \(\alpha=0.30\).  The pressure--displacement curves compare \(\beta=0\), \(0.15\), and the associated value \(\beta=\alpha\), with wall displacement reported as \(E u(a)/(ka)\).  The field plots use \(c/a=2.5\); the upper half of the polar map is associated and the lower half uses \(\beta=0.10\).

Together, the examples connect the directional formulation with active sets, hardening fields, pressure-sensitive stress states, and flow-dependent displacement fields.

\section{Viscous extension and the rate-independent limit}
\label{sec:ss_viscous}

Overstress and relaxation models can be interpreted as smooth kinetic
regularizations of a rate-independent graph \citep{Perzyna1966,DuvautLions1976,
BodnerPartom1975,OrtizStainier1999}. Here the viscous potential acts on the same
directional force that defines the rate-independent boundary; viscosity changes
the kinetics without redefining the stress gauge, resistance, or hardening
forces. The continuous rate functional is derived first. Its backward-Euler
residual belongs to the subsequent time-discrete formulation.

\subsection{A continuous rate functional}

All preceding rate-independent relations were written with the one-sided
variation $\delta\bm\lambda$.  Viscosity introduces a physical time scale, so
this section parameterizes the same paths by
$\delta\bm\lambda=\dot{\bm\lambda}\,\delta t$ and uses explicit rates.
Viscosity is introduced before time discretization.  At a fixed material state,
form the scalar rate functional
\begin{equation}
  \mathcal P_t[\dot{\bm u},\dot{\bm\lambda}]
  :=\frac{\dd}{\dd t}\Pi_t
  +\int_\Omega\mathcal V(\dot{\bm\lambda})\,\dd x,
  \qquad \dot\lambda_a\ge0,
  \label{eq:ss_vp_rate_functional}
\end{equation}
where every internal rate in \(\dot\Pi_t\) is restricted by the same tangent
relations that were used in the first variation.  After equilibrium has been
used, its internal part is
\begin{equation}
  \mathcal P_{t,\rm int}
  =\int_\Omega
  \left[-\bm F\cdot\dot{\bm\lambda}
  +\mathcal V(\dot{\bm\lambda})\right]\dd x.
  \label{eq:ss_vp_reduced_rate_functional}
\end{equation}
Variation with respect to an admissible rate gives
\begin{equation}
  \bm0\in-\bm F
  +\partial_{\dot{\bm\lambda}}\mathcal V
  +N_{\mathbb R_+^N}(\dot{\bm\lambda}).
  \label{eq:ss_vp_master}
\end{equation}
Thus the state functional still generates \(\bm F\), while the optional
viscous term changes the kinetic relation.  Equation \eqref{eq:ss_vp_master} is
a continuous-time statement.

For a positive rate and differentiable \(\mathcal V\),
\begin{equation}
  F_a=\frac{\partial\mathcal V}{\partial\dot\lambda_a}.
  \label{eq:ss_vp_overstress_general}
\end{equation}
With
\begin{equation}
  \mathcal V(\dot{\bm\lambda})
  =\sum_{a=1}^{N}
  \frac{\eta_a\dot\lambda_{0a}}{m_a+1}
  \left(\frac{\dot\lambda_a}{\dot\lambda_{0a}}\right)^{m_a+1},
  \qquad
  \eta_a>0,\quad\dot\lambda_{0a}>0,\quad m_a>0,
  \qquad \dot\lambda_a\ge0,
  \label{eq:ss_vp_power_potential}
\end{equation}
The corresponding activity rate is
\begin{equation}
  \boxed{
  \dot\lambda_a
  =\dot\lambda_{0a}
  \left\langle\frac{F_a}{\eta_a}\right\rangle_+^{1/m_a}.}
  \label{eq:ss_vp_perzyna}
\end{equation}
The resulting Perzyna-type law is the stationarity condition of the rate
functional driven by the previously defined directional force.  Its
force-space dual is
\begin{equation}
  \mathcal V_a^*(F_a)
  =\frac{m_a}{m_a+1}\eta_a\dot\lambda_{0a}
  \left\langle\frac{F_a}{\eta_a}\right\rangle_+^{(m_a+1)/m_a},
  \qquad
  \dot\lambda_a=\mathcal V_{a,F_a}^*.
  \label{eq:ss_vp_dual}
\end{equation}

\subsection{Associated, non-associated, and coupled viscosity}

The kinetic law changes the activity rate without altering the prescribed tangent geometry. Associated flow has $G_{,\bxi}\parallel F_{G,\bxi}$, whereas non-associated flow has nonparallel gradients. In both cases the energy variation gives $F_G=\bxi:G_{,\bxi}-R$ and the path relation gives $\dot\bepsp=G_{,\bxi}\dot\lambda$. A coupled potential
\begin{equation}
  \mathcal V
  =\frac12\dot{\bm\lambda}^{\T}\bm\eta\dot{\bm\lambda},
  \qquad \bm\eta=\bm\eta^{\T}\succ0,
  \label{eq:ss_vp_coupled_potential}
\end{equation}
produces
\begin{equation}
  \bm F=\bm\eta\dot{\bm\lambda}
  \quad\hbox{on the positive-rate set}.
  \label{eq:ss_vp_coupled_law}
\end{equation}
Off-diagonal entries of \(\bm\eta\) describe kinetic coupling, whereas
cross-derivatives of \(W(\bm\lambda)\) describe energetic hardening coupling.

\subsection{Dissipation and the rate-independent limit}

For
\(W=\sum_a\sigma_{0a}\lambda_a+H(\bm\lambda)\), the dissipation is
\begin{equation}
  \mathcal D
  =\sum_a\sigma_{0a}\dot\lambda_a
   +\sum_a\mathcal V_{,\dot\lambda_a}\dot\lambda_a
  \ge0.
  \label{eq:ss_vp_dissipation}
\end{equation}
As the viscosity scale tends to zero while active rates remain bounded,
\Cref{eq:ss_vp_overstress_general} enforces \(F_a\to0\) on active directions;
when a rate vanishes, the normal cone retains \(F_a\le0\).  The continuous
viscous graph therefore converges to the rate-independent KKT graph.

A backward-Euler discretization is given in Appendix~\ref{app:ss_discrete}.

\section{Discussion and conclusions}
\label{sec:discussion}
\label{sec:conclusions}

This paper has presented a directional-stationarity formulation of small-strain
plasticity in which one scalar functional yields the stress, internal-variable
forces, and activity resistance, while the one-sided tangent set specifies the
admissible plastic variations. Restricting the first variation to those paths
defines the directional forces and leads to
$F_a\le0$, $\delta\lambda_a\ge0$, and
$F_a\delta\lambda_a=0$. In several dimensions, the support direction of a
plastic metric gives associated normality. The self-similarity relation further
shows that a degree-one stress gauge generates a homogeneous family
$F=\sigma_*(\rho/\sigma_*)^m-R$, separating surface shape and flow direction
from isotropic scale and energetic translation.

Non-associated response is represented by a tangent that is not normal to the
generated boundary. Its use requires positive directional work, regularity of
the tangent map, and a stable active Jacobian; at corners or apexes the tangent
is naturally set-valued. Multiple mechanisms and hardening laws enter through
additional activity variables and the derivatives of $W$ and $\psi^p$.
Gradient terms modify the resistance through spatial derivatives, while a
viscous potential supplies rate dependence without changing the underlying
directional force.

The analytical solutions provide complementary checks of the formulation.
Multi-threshold torsion produces spatially separated active annuli, the
non-uniform bar gives an explicit kinematic-hardening field, the Hill-type
annulus recovers a pressure-sensitive limit state, and the cavity-expansion
solution separates strength from dilatancy through associated and
non-associated tangents. The appendices collect the time-discrete constitutive
integration and dimensional checks.

The most immediate extensions concern well-posedness for selected
non-associated and gradient models, consistent finite-element linearization,
and calibration against cyclic and rate-dependent experiments.

\section*{Acknowledgments}
The research is supported by the Fundamental Research Funds for the Central Universities (Grant No. 02302350113).

\section*{Data Availability Statement}
The data presented in this study are available on request from the corresponding author.

\section*{Conflicts of Interest}
The authors declare no conflicts of interest.

\appendix
\section{Time-discrete constitutive integration}
\label{app:ss_discrete}

This appendix collects the finite-increment formulas used for implementation. The continuum theory in the main text is written with \(\delta\lambda_a\); finite increments \(\Delta\lambda_a\) enter after time discretization.

\subsection{Radial return for the mixed-hardening example}
For a step from \(t_n\) to \(t_{n+1}\), define the finite numerical increment
\(\Delta p=p_{n+1}-p_n\).  The trial state is
\begin{equation}
  \bsig^{\rm tr}=\mathbb C:(\beps_{n+1}-\bepsp_n),
  \qquad
  \bxi^{\rm tr}=\dev\bsig^{\rm tr}-\bbeta_n,
  \qquad
  q^{\rm tr}=\sqrt{\frac32\bxi^{\rm tr}:\bxi^{\rm tr}},
  \label{eq:ss_example_trial}
\end{equation}
with
\begin{equation}
  F^{\rm tr}=q^{\rm tr}-\sigma_y-H_i p_n.
  \label{eq:ss_example_trial_F}
\end{equation}
The discrete KKT conditions are
\begin{equation}
  F_{n+1}\le0,
  \qquad
  \Delta p\ge0,
  \qquad
  F_{n+1}\Delta p=0.
\end{equation}
If \(F^{\rm tr}\le0\), \(\Delta p=0\).  On the active radial branch,
\begin{equation}
  q_{n+1}=q^{\rm tr}-(3G_s+C_k)\Delta p,
  \label{eq:ss_example_q_return}
\end{equation}
so
\begin{equation}
  \boxed{
  \Delta p
  =\frac{\langle q^{\rm tr}-\sigma_y-H_i p_n\rangle_+}
  {3G_s+C_k+H_i}.}
  \label{eq:ss_example_closed_return}
\end{equation}
The state update is
\begin{align}
  \dev\bsig_{n+1}
  &=\dev\bsig^{\rm tr}-2G_s\Delta p\,\bN,\\
  \bbeta_{n+1}
  &=\bbeta_n+\frac23C_k\Delta p\,\bN,\\
  p_{n+1}&=p_n+\Delta p.
  \label{eq:ss_example_state_update}
\end{align}
Here \(\Delta p\) denotes a finite time-discrete increment, as in the algorithms that follow.

\subsection{Backward-Euler residual for viscous evolution}
For a step of size \(\Delta t\), define
\(\Delta\lambda_a=\lambda_{a,n+1}-\lambda_{a,n}\).  Backward Euler applied to
\Cref{eq:ss_vp_perzyna} gives
\begin{equation}
  r_a^{\rm vp}
  =\Delta\lambda_a
  -\Delta t\,\dot\lambda_{0a}
  \left\langle\frac{F_{a,n+1}}{\eta_a}\right\rangle_+^{1/m_a}=0.
  \label{eq:ss_vp_discrete_residual}
\end{equation}
On a positive-overstress branch,
\begin{equation}
  \frac{\partial r_a^{\rm vp}}{\partial y}
  =\frac{\partial\Delta\lambda_a}{\partial y}
  -\frac{\Delta t\,\dot\lambda_{0a}}{m_a\eta_a}
  \left(\frac{F_{a,n+1}}{\eta_a}\right)^{1/m_a-1}
  \frac{\partial F_{a,n+1}}{\partial y}.
  \label{eq:ss_vp_residual_derivative}
\end{equation}
For the mixed-hardening example,
\begin{equation}
  r^{\rm vp}(\Delta p)
  =\Delta p-\Delta t\,\dot p_0
  \left\langle
  \frac{q^{\rm tr}-(3G_s+C_k)\Delta p
  -\sigma_y-H_i(p_n+\Delta p)}{\eta}
  \right\rangle_+^{1/m}=0.
  \label{eq:ss_vp_j2_residual}
\end{equation}
For \(m=1\),
\begin{equation}
  \boxed{
  \Delta p
  =\frac{(\Delta t\dot p_0/\eta)
  \langle q^{\rm tr}-\sigma_y-H_i p_n\rangle_+}
  {1+(\Delta t\dot p_0/\eta)(3G_s+C_k+H_i)}.}
  \label{eq:ss_vp_j2_closed}
\end{equation}
Here \(\Delta p\) and \(\Delta\lambda_a\) denote finite increments of the
time-discrete problem.

Return mapping and consistent linearization are mature tools of computational
plasticity \citep{OrtizSimo1986,SimoKennedyGovindjee1988,SimoHughes1998,
deSouzaNeto2008,DunnePetrinic2005}.  The continuum equalities and inequalities derived above provide the local
residual equations used by these algorithms.

\subsection{Active-set residuals}

For a strain increment from $t_n$ to $t_{n+1}$, let $\Delta\lambda_a\ge0$ be the direction increments. A backward-Euler update is
\begin{equation}
  \bm z_{I,n+1}
  =
  \bm z_{I,n}
  +\sum_{a\in\mathcal A}\Delta\lambda_a
  \mathcal M_{Ia,n+1}.
  \label{eq:BE_z}
\end{equation}
The local residuals for an assumed active set are
\begin{equation}
  \bm r_{z_I}
  =
  \bm z_{I,n+1}-\bm z_{I,n}
  -\sum_{a\in\mathcal A}\Delta\lambda_a\mathcal M_{Ia,n+1},
  \label{eq:r_z}
\end{equation}
\begin{equation}
  r_{F_a}=F_{a,n+1},
  \qquad a\in\mathcal A.
  \label{eq:r_F}
\end{equation}
Newton iteration solves \Cref{eq:r_z,eq:r_F}. The active set is then checked: active increments must remain nonnegative, and inactive directional forces must remain nonpositive. This is the direct computational counterpart of the one-sided tangent cone.

A compact complementarity residual can also be written with an active flag $A_a\in\{0,1\}$,
\begin{equation}
  r_{{\rm comp},a}
  =A_aF_a+(1-A_a)\Delta\lambda_a.
\end{equation}
Semi-smooth alternatives may replace the active flags with Fischer--Burmeister or min-function residuals.

\subsection{Block residual, discrete Jacobian, and active-set algorithm}

For implementation, collect all tensorial and scalar internal variables in the
state vector $\bm w$ and define the stacked direction vector $\bm d_a$ by
vectorizing the directions $\{\mathcal M_{Ia}\}_{I=1}^{M}$. For a fixed active set
$\mathcal A$, the local unknown and residual can be written as
\begin{equation}
  \bm y=
  \begin{bmatrix}
    \bm w_{n+1}\\[1mm]
    \Delta\bm\lambda_{\mathcal A}
  \end{bmatrix},
  \qquad
  \bm r(\bm y,\beps_{n+1})=
  \begin{bmatrix}
    \bm r_w\\[1mm]
    \bm F_{\mathcal A}
  \end{bmatrix}=\bm0,
  \label{eq:ss_compact_residual}
\end{equation}
where
\begin{equation}
  \bm r_w
  =\bm w_{n+1}-\bm w_n
  -\sum_{b\in\mathcal A}\Delta\lambda_b
  \bm d_b(\bm w_{n+1},\beps_{n+1}).
  \label{eq:ss_state_residual}
\end{equation}
If the directional force has no explicit multiplier dependence at fixed state, the
exact Newton matrix has the block form
\begin{equation}
  \bm J=
  \frac{\partial\bm r}{\partial\bm y}
  =
  \begin{bmatrix}
    \bm I-\displaystyle\sum_{b\in\mathcal A}\Delta\lambda_b
      \bm d_{b,w}
    & -\bm D_{\mathcal A}\\[2mm]
    \bm F_{\mathcal A,w} & \bm0
  \end{bmatrix},
  \qquad
  \bm D_{\mathcal A}
  =\begin{bmatrix}\bm d_{a_1}&\cdots&\bm d_{a_m}\end{bmatrix}.
  \label{eq:ss_block_jacobian}
\end{equation}
Any explicit multiplier dependence is included by replacing the lower-right
zero block with $\bm F_{\mathcal A,\Delta\lambda}$. Once the active set has
converged, exact differentiation of the discrete equations gives
\begin{equation}
  \frac{\dd\bm y}{\dd\beps_{n+1}}
  =-\bm J^{-1}\bm r_{,\beps},
  \qquad
  \mathbb C_{\rm alg}
  =\frac{\partial\bsig}{\partial\beps}
  +\frac{\partial\bsig}{\partial\bm y}:
   \frac{\dd\bm y}{\dd\beps_{n+1}}.
  \label{eq:ss_discrete_tangent}
\end{equation}
This differentiation is performed with the final active set held fixed; at a
switching point the constitutive map is only piecewise differentiable.

\begin{algorithm}[t]
\caption{Generic multimechanism backward-Euler active-set update}
\label{alg:ss_active_set}
\begin{algorithmic}[1]
\Require $\beps_{n+1}$, converged state $\bm w_n$, tolerances
\State Evaluate the elastic trial state and all trial directional forces $F_a^{\rm tr}$
\State Set $\mathcal A\gets\{a:F_a^{\rm tr}>0\}$ and initialize $\Delta\lambda_a\ge0$
\Repeat
  \Repeat
    \State Assemble $\bm r$ and the exact block Jacobian $\bm J$
    \State Solve $\bm J\,\delta\bm y=-\bm r$ and apply a safeguarded Newton update
  \Until{$\lVert\bm r\rVert$ satisfies the local tolerance}
  \State Remove active directions with $\Delta\lambda_a<0$
  \State Add inactive directions with $F_a>0$
\Until{all complementarity conditions are satisfied}
\State Evaluate $\bsig_{n+1}$ and $\mathbb C_{\rm alg}$ from \Cref{eq:ss_discrete_tangent}
\State \Return $\{\bsig_{n+1},\bm w_{n+1},\mathbb C_{\rm alg},\mathcal A\}$
\end{algorithmic}
\end{algorithm}
\Cref{alg:ss_active_set} summarizes the discrete active-set procedure; its
finite increments enter at this time-discrete stage.

\subsection{Continuum elastoplastic tangent}

The continuum tangent takes a simple form when directions and hardening moduli are linearized at the current state. Let
\begin{equation}
  \bm n_a=\frac{\partial F_a}{\partial\bsig},
  \qquad
  \bm m_a=\frac{\partial\bepsp}{\partial\lambda_a}
\end{equation}
be the yield normal and plastic-strain direction, respectively. For linear elasticity,
\begin{equation}
  \delta\bsig
  =\mathbb C:
  \left(
    \delta\beps-\sum_b\bm m_b\delta\lambda_b
  \right).
\end{equation}
Define the generalized hardening contribution as the total internal-state
derivative induced by mechanism $b$ at fixed stress,
\begin{equation}
  h_{ab}
  =-
  \left.
  \mathrm D_b^{\rm int}F_a
  \right|_{\bsig},
  \label{eq:hab}
\end{equation}
and the active-set matrix
\begin{equation}
  A_{ab}
  =\bm n_a:\mathbb C:\bm m_b+h_{ab}.
  \label{eq:Aab}
\end{equation}
Consistency $\delta F_a=0$ gives
\begin{equation}
  \delta\lambda_a
  =
  \sum_{b\in\mathcal A}
  [\bm A^{-1}]_{ab}
  (\bm n_b:\mathbb C):\delta\beps.
\end{equation}
The elastoplastic tangent is therefore
\begin{equation}
  \boxed{
  \mathbb C^{\rm ep}
  =
  \mathbb C
  -
  \sum_{a,b\in\mathcal A}
  (\mathbb C:\bm m_a)
  [\bm A^{-1}]_{ab}
  (\bm n_b:\mathbb C)
  }.
  \label{eq:Cep}
\end{equation}
For associated flow, $\bm m_a=\bm n_a$ and the tangent is symmetric when the hardening matrix is symmetric. For non-associated flow, the tangent is generally nonsymmetric. A fully algorithmic tangent follows by differentiating the converged discrete residuals through the implicit-function theorem; \Cref{eq:Cep} is the corresponding continuum form and a useful verification limit.

\subsection{Backward-Euler update for dynamic recovery}
For backward Euler integration,
\begin{equation}
  \balpha_{i,n+1}
  =
  \frac{
    \balpha_{i,n}+\frac23C_i\Delta\lambda\bN_{n+1}
  }{
    1+\gamma_i\Delta\lambda
  }.
  \label{eq:chaboche_update}
\end{equation}
The remaining scalar unknown is determined from $F_{n+1}=0$ together with the stress update.

\FloatBarrier

\section{Dimensional and coordinate checks}
\label{app:ss_dimensional_check}
\subsection{Dimensions and reparameterization}

Let $[\psi]$ denote energy per unit volume. For a general internal variable
$\bm z_I$, thermodynamic duality requires
\begin{equation}
  [\mathcal X_I][\bm z_I]=[\psi].
  \label{eq:ss_dual_dimensions}
\end{equation}
With dimensionless accumulated activity, $[\mathcal M_{Ia}]=[\bm z_I]$ and
therefore
\begin{equation}
  [\mathcal X_I\mathbin{\bullet}\mathcal M_{Ia}]
  =[R_a]=[F_a]=[\psi].
  \label{eq:ss_obstruction_dimensions}
\end{equation}
This check also distinguishes a stress-valued directional driving force from an
equivalent geometric representation of a yield locus.

The normalization of an activity variable is not unique. Let $c_a>0$ and define
\begin{equation}
  \widetilde\lambda_a=\frac{\lambda_a}{c_a},
  \qquad
  \widetilde{\mathcal M}_{Ia}=c_a\mathcal M_{Ia},
  \qquad
  \widetilde\sigma_{0a}=c_a\sigma_{0a}.
  \label{eq:ss_direction_rescaling}
\end{equation}
For $\bm C=\operatorname{diag}(c_1,\ldots,c_N)$, set
$\widetilde H(\widetilde{\bm\lambda})=H(\bm C\widetilde{\bm\lambda})$.
Then
\begin{equation}
  \delta{\bm z}_I
  =\sum_a\delta\lambda_a\mathcal M_{Ia}
  =\sum_a\delta\widetilde\lambda_a\widetilde{\mathcal M}_{Ia},
  \qquad
  \widetilde R_a=c_aR_a,
  \qquad
  \widetilde F_a=c_aF_a.
  \label{eq:ss_rescaling_invariance}
\end{equation}
The resistance contribution to the functional and the directional complementarity product are
unchanged:
\begin{equation}
  \widetilde R_a\,\dd\widetilde\lambda_a=R_a\,\dd\lambda_a,
  \qquad
  \widetilde F_a\,\delta\widetilde\lambda_a
  =F_a\,\delta\lambda_a.
  \label{eq:ss_rescaling_power}
\end{equation}
Thus direction scaling is a constitutive coordinate choice. Reporting the chosen
normalization remains essential because numerical multiplier values and
hardening parameters depend on it.

\AtNextBibliography{\small}
\printbibliography

@article{AbdelKarimOhno2000,
  author  = {Abdel-Karim, M. and Ohno, N.},
  title   = {Kinematic hardening model suitable for ratchetting with steady-state},
  journal = {International Journal of Plasticity},
  volume  = {16},
  number  = {3-4},
  pages   = {225--240},
  year    = {2000},
  doi     = {10.1016/S0749-6419(99)00052-2}
}

@article{Aifantis1984,
  author  = {Aifantis, E. C.},
  title   = {On the microstructural origin of certain inelastic models},
  journal = {Journal of Engineering Materials and Technology},
  volume  = {106},
  number  = {4},
  pages   = {326--330},
  year    = {1984},
  doi     = {10.1115/1.3225725}
}

@techreport{ArmstrongFrederick1966,
  author      = {Armstrong, P. J. and Frederick, C. O.},
  title       = {A mathematical representation of the multiaxial Bauschinger effect},
  institution = {Central Electricity Generating Board},
  number      = {RD/B/N 731},
  year        = {1966}
}

@article{Asaro1983,
  author  = {Asaro, R. J.},
  title   = {Micromechanics of crystals and polycrystals},
  journal = {Advances in Applied Mechanics},
  volume  = {23},
  pages   = {1--115},
  year    = {1983},
  doi     = {10.1016/S0065-2156(08)70242-4}
}

@article{Barlat1991,
  author  = {Barlat, F. and Lege, D. J. and Brem, J. C.},
  title   = {A six-component yield function for anisotropic materials},
  journal = {International Journal of Plasticity},
  volume  = {7},
  number  = {7},
  pages   = {693--712},
  year    = {1991},
  doi     = {10.1016/0749-6419(91)90033-Z}
}

@article{BodnerPartom1975,
  author  = {Bodner, S. R. and Partom, Y.},
  title   = {Constitutive equations for elastic-viscoplastic strain-hardening materials},
  journal = {Journal of Applied Mechanics},
  volume  = {42},
  number  = {2},
  pages   = {385--389},
  year    = {1975},
  doi     = {10.1115/1.3423586}
}

@article{CazacuBarlat2004,
  author  = {Cazacu, O. and Barlat, F.},
  title   = {A criterion for description of anisotropy and yield differential effects in pressure-insensitive metals},
  journal = {International Journal of Plasticity},
  volume  = {20},
  number  = {11},
  pages   = {2027--2045},
  year    = {2004},
  doi     = {10.1016/j.ijplas.2003.11.013}
}

@article{Chaboche1986,
  author  = {Chaboche, J. L.},
  title   = {Time-independent constitutive theories for cyclic plasticity},
  journal = {International Journal of Plasticity},
  volume  = {2},
  number  = {2},
  pages   = {149--188},
  year    = {1986},
  doi     = {10.1016/0749-6419(86)90010-0}
}

@article{Chaboche1989,
  author  = {Chaboche, J. L.},
  title   = {Constitutive Equations for Cyclic Plasticity and Cyclic Viscoplasticity},
  journal = {International Journal of Plasticity},
  volume  = {5},
  number  = {3},
  pages   = {247--302},
  year    = {1989},
  doi     = {10.1016/0749-6419(89)90015-6}
}

@article{Chaboche2008,
  author  = {Chaboche, J. L.},
  title   = {A review of some plasticity and viscoplasticity constitutive theories},
  journal = {International Journal of Plasticity},
  volume  = {24},
  number  = {10},
  pages   = {1642--1693},
  year    = {2008},
  doi     = {10.1016/j.ijplas.2008.03.009}
}

@article{ColemanNoll1963,
  author  = {Coleman, B. D. and Noll, W.},
  title   = {The thermodynamics of elastic materials with heat conduction and viscosity},
  journal = {Archive for Rational Mechanics and Analysis},
  volume  = {13},
  pages   = {167--178},
  year    = {1963},
  doi     = {10.1007/BF01262690}
}

@article{CollinsHoulsby1997,
  author  = {Collins, I. F. and Houlsby, G. T.},
  title   = {Application of thermomechanical principles to the modelling of geotechnical materials},
  journal = {Proceedings of the Royal Society of London. Series A: Mathematical, Physical and Engineering Sciences},
  volume  = {453},
  number  = {1964},
  pages   = {1975--2001},
  year    = {1997},
  doi     = {10.1098/rspa.1997.0107}
}

@article{DafaliasPopov1976,
  author  = {Dafalias, Y. F. and Popov, E. P.},
  title   = {Plastic internal variables formalism of cyclic plasticity},
  journal = {Journal of Applied Mechanics},
  volume  = {43},
  number  = {4},
  pages   = {645--651},
  year    = {1976},
  doi     = {10.1115/1.3423948}
}

@book{deSouzaNeto2008,
  author    = {{de Souza Neto}, E. A. and Peri{\'c}, D. and Owen, D. R. J.},
  title     = {Computational Methods for Plasticity: Theory and Applications},
  publisher = {Wiley},
  address   = {Chichester},
  year      = {2008},
  doi       = {10.1002/9780470694626}
}

@article{Drucker1959,
  author  = {Drucker, D. C.},
  title   = {A definition of stable inelastic material},
  journal = {Journal of Applied Mechanics},
  volume  = {26},
  number  = {1},
  pages   = {101--106},
  year    = {1959},
  doi     = {10.1115/1.4011929}
}

@article{DruckerPrager1952,
  author  = {Drucker, D. C. and Prager, W.},
  title   = {Soil mechanics and plastic analysis or limit design},
  journal = {Quarterly of Applied Mathematics},
  volume  = {10},
  number  = {2},
  pages   = {157--165},
  year    = {1952},
  doi     = {10.1090/qam/48291}
}

@book{DunnePetrinic2005,
  author    = {Dunne, F. and Petrinic, N.},
  title     = {Introduction to Computational Plasticity},
  publisher = {Oxford University Press},
  address   = {Oxford},
  year      = {2005}
}

@book{DuvautLions1976,
  author    = {Duvaut, G. and Lions, J.-L.},
  title     = {Inequalities in Mechanics and Physics},
  publisher = {Springer},
  address   = {Berlin},
  year      = {1976}
}

@incollection{FleckHutchinson1997,
  author    = {Fleck, N. A. and Hutchinson, J. W.},
  title     = {Strain gradient plasticity},
  booktitle = {Advances in Applied Mechanics},
  editor    = {Hutchinson, J. W. and Wu, T. Y.},
  volume    = {33},
  pages     = {295--361},
  publisher = {Academic Press},
  year      = {1997},
  doi       = {10.1016/S0065-2156(08)70388-0}
}

@article{FleckHutchinson2001,
  author  = {Fleck, N. A. and Hutchinson, J. W.},
  title   = {A reformulation of strain gradient plasticity},
  journal = {Journal of the Mechanics and Physics of Solids},
  volume  = {49},
  number  = {10},
  pages   = {2245--2271},
  year    = {2001},
  doi     = {10.1016/S0022-5096(01)00049-7}
}

@article{FleckWillis2009,
  author  = {Fleck, N. A. and Willis, J. R.},
  title   = {A mathematical basis for strain-gradient plasticity theory. Part I: Scalar plastic multiplier},
  journal = {Journal of the Mechanics and Physics of Solids},
  volume  = {57},
  number  = {1},
  pages   = {161--177},
  year    = {2009},
  doi     = {10.1016/j.jmps.2008.09.010}
}

@article{Gurtin2000,
  author  = {Gurtin, M. E.},
  title   = {On the plasticity of single crystals: Free energy, microforces, plastic-strain gradients},
  journal = {Journal of the Mechanics and Physics of Solids},
  volume  = {48},
  number  = {5},
  pages   = {989--1036},
  year    = {2000},
  doi     = {10.1016/S0022-5096(99)00059-9}
}

@article{GurtinAnand2005,
  author  = {Gurtin, M. E. and Anand, L.},
  title   = {A theory of strain-gradient plasticity for isotropic, plastically irrotational materials. Part I: Small deformations},
  journal = {Journal of the Mechanics and Physics of Solids},
  volume  = {53},
  number  = {7},
  pages   = {1624--1649},
  year    = {2005},
  doi     = {10.1016/j.jmps.2004.12.008}
}

@article{GurtinAnand2009,
  author  = {Gurtin, M. E. and Anand, L.},
  title   = {Thermodynamics applied to gradient theories involving the accumulated plastic strain: The theories of Aifantis and Fleck and Hutchinson and their generalization},
  journal = {Journal of the Mechanics and Physics of Solids},
  volume  = {57},
  number  = {3},
  pages   = {405--421},
  year    = {2009},
  doi     = {10.1016/j.jmps.2008.12.002}
}

@article{HalphenNguyen1975,
  author  = {Halphen, Bernard and Nguyen, Quoc Son},
  title   = {Sur les mat{\'e}riaux standards g{\'e}n{\'e}ralis{\'e}s},
  journal = {Journal de M{\'e}canique},
  volume  = {14},
  number  = {1},
  pages   = {39--63},
  year    = {1975}
}

@article{Hencky1924,
  author  = {Hencky, H.},
  title   = {Zur Theorie plastischer Deformationen und der hierdurch im Material hervorgerufenen Nachspannungen},
  journal = {Zeitschrift f{\"u}r Angewandte Mathematik und Mechanik},
  volume  = {4},
  number  = {4},
  pages   = {323--334},
  year    = {1924}
}

@article{Hill1948Anisotropic,
  author  = {Hill, R.},
  title   = {A theory of the yielding and plastic flow of anisotropic metals},
  journal = {Proceedings of the Royal Society of London. Series A},
  volume  = {193},
  number  = {1033},
  pages   = {281--297},
  year    = {1948},
  doi     = {10.1098/rspa.1948.0045}
}

@book{Hill1950,
  author    = {Hill, R.},
  title     = {The Mathematical Theory of Plasticity},
  publisher = {Clarendon Press},
  address   = {Oxford},
  year      = {1950}
}

@article{Hosford1972,
  author  = {Hosford, W. F.},
  title   = {A generalized isotropic yield criterion},
  journal = {Journal of Applied Mechanics},
  volume  = {39},
  number  = {2},
  pages   = {607--609},
  year    = {1972},
  doi     = {10.1115/1.3422732}
}

@book{HoulsbyPuzrin2007,
  author    = {Houlsby, G. T. and Puzrin, A. M.},
  title     = {Principles of Hyperplasticity: An Approach to Plasticity Theory Based on Thermodynamic Principles},
  publisher = {Springer},
  address   = {London},
  year      = {2007},
  doi       = {10.1007/978-1-84628-240-9}
}

@article{Huber1904,
  author  = {Huber, M. T.},
  title   = {W{\l}a{\'s}ciwa praca odkszta{\l}cenia jako miara wyt{\k e}{\.{z}}enia materya{\l}u},
  journal = {Czasopismo Techniczne},
  volume  = {22},
  pages   = {38--50},
  year    = {1904}
}

@article{Iwan1967,
  author  = {Iwan, W. D.},
  title   = {On the formulation and numerical solution of a class of multiaxial plasticity problems},
  journal = {Journal of Applied Mechanics},
  volume  = {34},
  number  = {3},
  pages   = {612--617},
  year    = {1967},
  doi     = {10.1115/1.3607754}
}

@article{Koiter1953,
  author  = {Koiter, W. T.},
  title   = {Stress-strain relations, uniqueness and variational theorems for elastic-plastic materials with a singular yield surface},
  journal = {Quarterly of Applied Mathematics},
  volume  = {11},
  pages   = {350--354},
  year    = {1953}
}

@book{Lubliner1990,
  author    = {Lubliner, J.},
  title     = {Plasticity Theory},
  publisher = {Macmillan},
  address   = {New York},
  year      = {1990}
}

@article{Mandel1973,
  author  = {Mandel, J.},
  title   = {Equations constitutives et directeurs dans les milieux plastiques et viscoplastiques},
  journal = {International Journal of Solids and Structures},
  volume  = {9},
  number  = {6},
  pages   = {725--740},
  year    = {1973},
  doi     = {10.1016/0020-7683(73)90114-6}
}

@article{Moreau1970,
  author  = {Moreau, J. J.},
  title   = {Sur les lois de frottement, de plasticit{\'e} et de viscosit{\'e}},
  journal = {Comptes Rendus de l'Acad{\'e}mie des Sciences, S{\'e}rie A},
  volume  = {271},
  pages   = {608--611},
  year    = {1970}
}

@article{Mroz1967,
  author  = {Mr{\'o}z, Z.},
  title   = {On the description of anisotropic workhardening},
  journal = {Journal of the Mechanics and Physics of Solids},
  volume  = {15},
  number  = {3},
  pages   = {163--175},
  year    = {1967},
  doi     = {10.1016/0022-5096(67)90030-0}
}

@article{OhnoWang1993,
  author  = {Ohno, N. and Wang, J. D.},
  title   = {Kinematic hardening rules with critical state of dynamic recovery, part I: Formulation and basic features for ratchetting behavior},
  journal = {International Journal of Plasticity},
  volume  = {9},
  number  = {3},
  pages   = {375--390},
  year    = {1993},
  doi     = {10.1016/0749-6419(93)90042-O}
}

@article{OrtizSimo1986,
  author  = {Ortiz, M. and Simo, J. C.},
  title   = {An analysis of a new class of integration algorithms for elastoplastic constitutive relations},
  journal = {International Journal for Numerical Methods in Engineering},
  volume  = {23},
  number  = {3},
  pages   = {353--366},
  year    = {1986},
  doi     = {10.1002/nme.1620230303}
}

@article{OrtizStainier1999,
  author  = {Ortiz, M. and Stainier, L.},
  title   = {The variational formulation of viscoplastic constitutive updates},
  journal = {Computer Methods in Applied Mechanics and Engineering},
  volume  = {171},
  number  = {3-4},
  pages   = {419--444},
  year    = {1999},
  doi     = {10.1016/S0045-7825(98)00219-9}
}

@article{PeirceAsaroNeedleman1983,
  author  = {Peirce, D. and Asaro, R. J. and Needleman, A.},
  title   = {Material rate dependence and localized deformation in crystalline solids},
  journal = {Acta Metallurgica},
  volume  = {31},
  number  = {12},
  pages   = {1951--1976},
  year    = {1983},
  doi     = {10.1016/0001-6160(83)90014-7}
}

@incollection{Perzyna1966,
  author    = {Perzyna, P.},
  title     = {Fundamental problems in viscoplasticity},
  booktitle = {Advances in Applied Mechanics},
  volume    = {9},
  pages     = {243--377},
  publisher = {Academic Press},
  year      = {1966},
  doi       = {10.1016/S0065-2156(08)70009-7}
}

@article{Prager1956,
  author  = {Prager, W.},
  title   = {A new method of analyzing stresses and strains in work-hardening plastic solids},
  journal = {Journal of Applied Mechanics},
  volume  = {23},
  pages   = {493--496},
  year    = {1956}
}

@article{Prandtl1924,
  author  = {Prandtl, L.},
  title   = {Spannungsverteilung in plastischen K{\"o}rpern},
  journal = {Proceedings of the First International Congress for Applied Mechanics},
  pages   = {43--54},
  year    = {1924}
}

@article{Reuss1930,
  author  = {Reuss, E.},
  title   = {Ber{\"u}cksichtigung der elastischen Form{\"a}nderung in der Plastizit{\"a}tstheorie},
  journal = {Zeitschrift f{\"u}r Angewandte Mathematik und Mechanik},
  volume  = {10},
  number  = {3},
  pages   = {266--274},
  year    = {1930}
}

@article{Rice1971,
  author  = {Rice, J. R.},
  title   = {Inelastic constitutive relations for solids: An internal-variable theory and its application to metal plasticity},
  journal = {Journal of the Mechanics and Physics of Solids},
  volume  = {19},
  number  = {6},
  pages   = {433--455},
  year    = {1971},
  doi     = {10.1016/0022-5096(71)90010-X}
}

@book{SimoHughes1998,
  author    = {Simo, J. C. and Hughes, T. J. R.},
  title     = {Computational Inelasticity},
  series    = {Interdisciplinary Applied Mathematics},
  volume    = {7},
  publisher = {Springer},
  address   = {New York},
  year      = {1998},
  doi       = {10.1007/b98904}
}

@article{SimoKennedyGovindjee1988,
  author  = {Simo, J. C. and Kennedy, J. G. and Govindjee, S.},
  title   = {Non-smooth multisurface plasticity and viscoplasticity. Loading/unloading conditions and numerical algorithms},
  journal = {International Journal for Numerical Methods in Engineering},
  volume  = {26},
  number  = {10},
  pages   = {2161--2185},
  year    = {1988},
  doi     = {10.1002/nme.1620261003}
}

@article{Tresca1864,
  author  = {Tresca, H.},
  title   = {M{\'e}moire sur l'{\'e}coulement des corps solides soumis {\`a} de fortes pressions},
  journal = {Comptes Rendus de l'Acad{\'e}mie des Sciences},
  volume  = {59},
  pages   = {754--758},
  year    = {1864}
}

@article{Ulloa2021,
  author  = {Ulloa, J. and Alessi, R. and Wambacq, J. and Degrande, G. and Fran{\c c}ois, S.},
  title   = {On the variational modeling of non-associative plasticity},
  journal = {International Journal of Solids and Structures},
  volume  = {217--218},
  pages   = {272--296},
  year    = {2021},
  doi     = {10.1016/j.ijsolstr.2020.10.027}
}

@article{vonMises1913,
  author  = {{von Mises}, R.},
  title   = {Mechanik der festen K{\"o}rper im plastisch-deformablen Zustand},
  journal = {Nachrichten von der Gesellschaft der Wissenschaften zu G{\"o}ttingen, Mathematisch-Physikalische Klasse},
  pages   = {582--592},
  year    = {1913}
}

@article{Hackl1997,
  author  = {Hackl, K.},
  title   = {Generalized standard media and variational principles in classical and finite strain elastoplasticity},
  journal = {Journal of the Mechanics and Physics of Solids},
  volume  = {45},
  number  = {5},
  pages   = {667--688},
  year    = {1997},
  doi     = {10.1016/S0022-5096(96)00110-X}
}

@article{HacklFischer2008,
  author  = {Hackl, K. and Fischer, F. D.},
  title   = {On the relation between the principle of maximum dissipation and inelastic evolution given by dissipation potentials},
  journal = {Proceedings of the Royal Society A},
  volume  = {464},
  number  = {2089},
  pages   = {117--132},
  year    = {2008},
  doi     = {10.1098/rspa.2007.0086}
}

@article{TeichtmeisterKeip2022,
  author  = {Teichtmeister, S. and Keip, M.-A.},
  title   = {A variational framework for the thermomechanics of gradient-extended dissipative solids---with applications to diffusion, damage and plasticity},
  journal = {Journal of Elasticity},
  volume  = {148},
  number  = {1},
  pages   = {81--126},
  year    = {2022},
  doi     = {10.1007/s10659-022-09884-6}
}

@article{Sewell1973a,
  author  = {Sewell, M. J.},
  title   = {The Governing Equations and Extremum Principles of Elasticity and Plasticity Generated from a Single Functional. Part I},
  journal = {Journal of Structural Mechanics},
  volume  = {2},
  number  = {1},
  pages   = {1--32},
  year    = {1973},
  doi     = {10.1080/03601217308905298}
}

@article{Sewell1973b,
  author  = {Sewell, M. J.},
  title   = {The Governing Equations and Extremum Principles of Elasticity and Plasticity Generated from a Single Functional. Part II},
  journal = {Journal of Structural Mechanics},
  volume  = {2},
  number  = {2},
  pages   = {135--158},
  year    = {1973},
  doi     = {10.1080/03601217308905303}
}

@article{DelPiero2018Variational,
  author  = {Del Piero, Gianpietro},
  title   = {The Variational Structure of Classical Plasticity},
  journal = {Mathematics and Mechanics of Complex Systems},
  volume  = {6},
  number  = {3},
  pages   = {137--180},
  year    = {2018},
  doi     = {10.2140/memocs.2018.6.137}
}

@article{AbatourForest2024Saturating,
  author  = {Abatour, Mohamed and Forest, Samuel},
  title   = {Strain Gradient Plasticity Based on Saturating Variables},
  journal = {European Journal of Mechanics - A/Solids},
  volume  = {104},
  pages   = {105016},
  year    = {2024},
  doi     = {10.1016/j.euromechsol.2023.105016}
}

@article{MukherjeeBanerjee2024ElasticGap,
  author  = {Mukherjee, Anjan and Banerjee, Biswanath},
  title   = {Elastic-Gap Free Formulation in Strain Gradient Plasticity Theory},
  journal = {Journal of Applied Mechanics},
  volume  = {91},
  number  = {6},
  pages   = {061008},
  year    = {2024},
  doi     = {10.1115/1.4064790}
}

@article{FuhgEtAl2025DataDriven,
  author  = {Fuhg, Jan N. and Padmanabha, Govinda Anantha and Bouklas, Nikolaos and Bahmani, Bahador and Sun, WaiChing and Vlassis, Nikolaos N. and Flaschel, Moritz and Carrara, Pietro and De Lorenzis, Laura},
  title   = {A Review on Data-Driven Constitutive Laws for Solids},
  journal = {Archives of Computational Methods in Engineering},
  volume  = {32},
  number  = {3},
  pages   = {1841--1883},
  year    = {2025},
  doi     = {10.1007/s11831-024-10196-2}
}

@article{CarterBookerYeung1986,
  author  = {Carter, J. P. and Booker, J. R. and Yeung, S. K.},
  title   = {Cavity Expansion in Cohesive Frictional Soils},
  journal = {G{\'e}otechnique},
  volume  = {36},
  number  = {3},
  pages   = {349--358},
  year    = {1986},
  doi     = {10.1680/geot.1986.36.3.349}
}

@article{YuHoulsby1991,
  author  = {Yu, Hai-Sui and Houlsby, Guy T.},
  title   = {Finite Cavity Expansion in Dilatant Soils: Loading Analysis},
  journal = {G{\'e}otechnique},
  volume  = {41},
  number  = {2},
  pages   = {173--183},
  year    = {1991},
  doi     = {10.1680/geot.1991.41.2.173}
}

@article{YuCarter2002,
  author  = {Yu, Hai-Sui and Carter, John P.},
  title   = {Rigorous Similarity Solutions for Cavity Expansion in Cohesive-Frictional Soils},
  journal = {International Journal of Geomechanics},
  volume  = {2},
  number  = {2},
  pages   = {233--258},
  year    = {2002},
  doi     = {10.1061/(ASCE)1532-3641(2002)2:2(233)}
}

@article{ChenWang2024Cavity,
  author  = {Chen, Sheng-Li and Wang, Xu},
  title   = {A Graphical Method for Undrained Analysis of Cavity Expansion in {Mohr--Coulomb} Soil},
  journal = {G{\'e}otechnique},
  volume  = {74},
  number  = {12},
  pages   = {1228--1240},
  year    = {2024},
  doi     = {10.1680/jgeot.22.00088}
}

\end{document}